\documentclass[reprint,
onecolumn,
amsmath,amssymb,
aps,
]{revtex4-2}

\usepackage{graphicx}
\usepackage{dcolumn}
\usepackage{bm}

\usepackage{xcolor}
\usepackage{xfrac}
\usepackage{float}
\usepackage{tabularx}
\usepackage{hyphenat}
\usepackage{makecell}
\usepackage{bbm}
\usepackage[T1]{fontenc}
\usepackage[utf8]{inputenc}
\usepackage[margin=0.8in]{geometry}

\usepackage{amsmath}
\graphicspath{ {./Images/} }
\usepackage{subfig}
\usepackage{float}
\usepackage[maxfloats=100]{morefloats}
\usepackage{multirow}
\usepackage{caption}
\usepackage{array}

\begin{document}

\preprint{APS/123-QED}

\title{ConvNet-Based Prediction of Droplet Collision Dynamics in Microchannels}

\author{S M Abdullah Al Mamun}
\author{Samaneh Farokhirad}%
 \email{samaneh.farokhirad@njit.edu}
\affiliation{Department of Mechanical and Industrial Engineering, New Jersey Institute of Technology, Newark, NJ 07114
}





\begin{abstract}

The dynamics of droplet collisions in microchannels are inherently complex, governed by multiple interdependent physical and geometric factors. Understanding and predicting the outcomes of these collisions—whether coalescence, reverse-back, or pass-over- pose significant challenges, particularly due to the deformability of droplets and the influence of key parameters such as viscosity ratios, density ratios, confinement, and initial offset of droplets. Traditional methods for analyzing these collisions, including computational simulations and experimental techniques, are time-consuming and resource-intensive, limiting their scalability for real-time applications. In this work, we explore a novel data-driven approach to predict droplet collision outcomes using convolutional neural networks (CNNs). The CNN-based approach presents a significant advantage over traditional methods, offering faster, scalable solutions for analyzing large datasets with varying physical parameters. Using a lattice Boltzmann method based on Cahn-Hilliard diffuse interface theory for binary immiscible fluids, we numerically generated droplet collision data under confined shear flow. This data, represented as droplet shapes, serves as input to the CNN model, which automatically learns hierarchical features from the images, allowing for accurate and efficient collision outcome predictions based on deformation and orientation. The model achieves a prediction accuracy of $0.972$, even on test datasets with varied density and viscosity ratios not included in training. Our findings suggest that the CNN-based models offer improved accuracy in predicting collision outcomes while drastically reducing computational and time constraints. This work highlights the potential of machine learning to advance droplet dynamics studies, providing a valuable tool for researchers in fluid dynamics and soft matter.

\end{abstract}

\maketitle


\section{\label{sec:level1}Introduction}

The study of droplet dynamics, particularly in multiphase flows, has attracted considerable attention in both academic research and industrial applications due to its broad relevance in fields such as targeted drug delivery~\cite{patra2018nano,zhao2013multiphase, singh2017nanoemulsion, sarker2005engineering}, enhanced oil recovery (EOR)~\cite{wasan1979role,li2020advances,son2014enhanced, jia2021potential}, microfluidics~\cite{xu1999gas, zhao2001co, zhao2011microfluidic, halliday2006improved}, cosmetics~\cite{bernardo2005integrated, patravale2008novel, kouhi2020edible}, emulsification~\cite{nilsson2006adsorption, lobo2003coalescence, jafari2008re}, materials processing~\cite{lee2008double, zhao2012bio}, chemical reactions~\cite{chen2011reactions, utada2005monodisperse}, biology engineering~\cite{abkarian2007swinging, fedosov2010multiscale, pivkin2008accurate} and many more. Droplet collisions, as a subtopic of this field, present an especially complex challenge due to the intricate interactions between droplets and the surrounding fluid. These interactions are governed by numerous physical and geometric parameters that affect droplet deformation, coalescence, and breakup, making accurate predictions of collision outcomes a difficult task. The complexity is further compounded in microfluidic systems, where confinement plays a critical role in altering the behavior of droplets during collisions. Our research focuses on predicting the outcome of droplet pair collisions in confined microchannel flow, using convolutional neural networks (CNN) by analyzing the shape of droplets during the collision process.

The motivation for this study stems from the need to overcome the limitations of traditional approaches in predicting droplet collision outcomes. In soft matter systems, such as emulsions, droplets are suspended in a continuous fluid and frequently undergo collisions. These collisions result in a range of outcomes—coalescence, reverse-back, or pass-over depending on the interplay of several parameters such as droplet size, velocity, viscosity ratio, density ratio, confinement, and initial offset. A deeper understanding of these collision outcomes is essential for optimizing processes in industries. However, predicting these outcomes is a non-trivial task due to the nonlinear nature of the governing factors and the sensitivity of collision behavior to small variations in system parameters.

The traditional methods for predicting droplet collision outcomes typically involve either phenomenological models or direct numerical simulations (DNS). Phenomenological models are often derived from simplified physical principles, where key parameters such as Weber number, Reynolds number, and Ohnesorge number are used to define collision regimes. One of the earliest models in this area was proposed by Ashgriz and Poo~\cite{ashgriz1990coalescence}, who classified droplet collisions into different modes (e.g., bouncing, coalescence, and reflexive or stretching breakup) based on a set of dimensionless numbers. Subsequent models by Jiang et al.~\cite{Jiang_Umemura_Law_1992} and Qian and Law~\cite{qian1997regimes} further refined these classifications, developing “collision maps” to predict outcomes based on experimental data. However, these models have limited applicability because they often require calibration with experimental data and cannot generalize to new parameter spaces. They also tend to break down when additional complexities like surface tension variations, viscosity contrasts, or confinement effects are introduced. 
In another study, Agarwal et al.~\cite{agarwal2019computational} investigated the collisions in fuel spray systems and used phenomenological models to predict the outcomes of binary droplet collisions. Their model achieved a prediction accuracy of around 64\%, highlighting the limitations of using such approaches for complex multiphase systems. Sommerfeld and Kuschel~\cite{sommerfeld2016modelling} extended this work by incorporating additional data points, including different fluid viscosities and droplet velocities, but observed a drop in accuracy to about 43\% when these variables were included. This suggests that traditional empirical correlations struggle to capture the complex physics involved, especially when dealing with highly deformable droplets in confined geometries. The use of simplified governing equations makes it difficult for these models to account for the full range of physical phenomena that occur during droplet collisions.

On the other hand, direct numerical simulations (DNS) have been used to model droplet collisions with higher accuracy, as they solve the full Navier-Stokes equations without relying on simplifying assumptions. Studies by Bianchi et al.~\cite{bianchi2015numerical} and Liu et al.~\cite{liu2018direct} have shown that DNS can predict collision outcomes with a high degree of accuracy. However, DNS is computationally expensive, particularly when simulating large systems or conducting parametric studies involving multiple variables. This makes DNS impractical for real-time applications or for generating large datasets required for building comprehensive collision maps. Moreover, even high-fidelity simulations may struggle to capture the effects of local geometry and deformation in microfluidic systems where confinement becomes a dominant factor.

Given the limitations of both phenomenological models and DNS, there is a clear need for alternative approaches that can provide accurate predictions without the computational burden of full-scale simulations. The advent of machine learning (ML) and, more specifically, deep learning techniques, offers a promising solution to this problem.

In recent years, ML has emerged as a powerful tool for modeling complex fluid dynamics phenomena. ML techniques are capable of identifying patterns in large datasets and developing predictive models that generalize well to new data. One of the key advantages of ML in fluid dynamics is its ability to model nonlinear relationships between variables, which is particularly useful when dealing with phenomena like turbulence, multiphase flows, or droplet collisions. Numerous studies have demonstrated the potential of ML in areas such as turbulence modeling, where traditional Reynolds-Averaged Navier-Stokes (RANS) or Large Eddy Simulations are computationally expensive and often inaccurate in complex flows.
For example, Duraisamy et al.~\cite{duraisamy2019turbulence} applied ML to develop closure models for turbulence, achieving better accuracy than traditional RANS models. Similarly, Ling et al.~\cite{ling2016reynolds} used deep learning to predict turbulence quantities from sparse data, showing that ML can improve both accuracy and efficiency in fluid flow predictions. 

On the other hand, reduced order modeling (ROM) technique plays a pivotal role in advancing fluid flow analysis and control~\cite{ghoreyshi2014reduced, lucia2004reduced, wang2018model}. ROM aims to simplify complex, high-dimensional fluid flow problems by creating low-dimensional models that retain essential dynamics. In fluid flow applications, data-driven sparse reconstruction also plays useful role when dealing with limited sensor data ~\cite{al2018extreme, jayaraman2019interplay, jayaraman2020data, al2020data, callaham2019robust}. For instances, in aerodynamic design, climate modeling, and industrial fluid processes, sparse reconstruction can provide detailed insights into the flow behavior around complex geometries using minimal measurements. 

These successes in data-driven modeling have inspired researchers to apply ML to other areas of fluid dynamics such as wall-impact breakage modeling in particle-laden flows~\cite{khalifa2022neural}. The ability of ML algorithms to automatically learn hierarchical features from data makes them particularly well-suited for problems where visual information, such as droplet deformation, plays a key role in determining the outcome.

Convolutional neural networks (CNNs), a subset of deep learning, have been widely applied in computer vision tasks due to their ability to learn spatial hierarchies of features from images. In the context of fluid dynamics, CNNs have been used to extract features from flow fields and predict outcomes such as vortex formation or separation in aerodynamic flows. For example, Guo et al.~\cite{guo2016convolutional} employed CNNs to predict flow fields around airfoils based on input geometric parameters, achieving high accuracy in their predictions. The success of CNNs in identifying patterns in visual data has opened up new possibilities for their application in multiphase flow problems, where the shape and deformation of objects like droplets are critical in determining behavior.

While CNNs have been extensively used in object recognition and image classification tasks in computer vision ~\cite{krizhevsky2012imagenet, simonyan2015very, he2016deep}, their application to droplet collision modeling is relatively new. The ability of CNNs to recognize and classify complex shapes makes them ideal for studying droplet interactions, where the deformation of the droplet can reveal key information about the forces acting on it during a collision. For instance, Jian et al.~\cite{khor2019using} utilized a convolutional
autoencoder model to classify droplet break-up moving through a constriction, highlighting the potential of CNNs in fluid dynamics applications.

In the case of droplet collisions, the shape of the droplets during the interaction reflects the balance of forces acting on them, including surface tension, viscous forces, and inertia. By using CNNs to analyze these shapes, we can predict the outcome of the collision whether the droplets will coalesce, reverse back, or pass over each other. This approach offers a significant advantage over traditional methods, as it allows for real-time predictions without the need for expensive simulations or experiments.

In this study, we propose a novel approach for predicting droplet collision outcomes using CNNs to analyze the shape of droplets during their interaction in confined microchannel flow. By training the CNN on a reasonably large dataset of simulated droplet collisions, we aim to develop a model that can accurately predict whether two droplets will coalesce, reverse back, or pass over each other based on their shape at the moment of collision. This approach leverages the ability of CNNs to extract hierarchical features from image data, capturing the subtle deformations that are indicative of different collision outcomes.
Unlike traditional methods, our approach does not rely on pre-defined empirical correlations or simplifying assumptions about the governing equations. Instead, it uses the raw image data to learn the complex relationships between droplet shape, orientation, and collision outcome. This allows our model to generalize across a wide range of parameters, including varying fluid viscosities, densities, droplet sizes, initial offset, and confinement geometries. Moreover, the use of CNNs allows for real-time predictions, making this approach suitable for applications where rapid decision-making is critical, such as in microfluidic device design or process control in industrial applications.

As is the case, the prediction of droplet collision outcomes in microfluidic systems presents a significant challenge due to the complexity of the interactions and the multitude of influencing parameters. Traditional methods, such as phenomenological models and DNS, face limitations in terms of accuracy, scalability, and computational cost. ML, and specifically CNNs, offers a promising alternative by leveraging the power of data-driven modeling to predict collision outcomes based on droplet shape. This study aims to contribute to the growing body of literature on ML in fluid dynamics by demonstrating the effectiveness of CNNs in predicting droplet collision outcomes in confined flows. First, simulations are performed under varied input parameters using a free-energy-based lattice Boltzmann method. The dataset of collision dynamics between droplet pairs is then collected and preprocessed to train a CNN-based predictive model capable of accurately classifying droplet collision outcomes based on shape analysis. The performance
of the CNN model is assessed using a broader range of simulation conditions beyond the training and validation datasets to evaluate its applicability and reliability in real-world scenarios. These conditions, in the context of droplet collision prediction, include variations in the density and viscosity of immiscible fluids, the initial offset distance between the two droplets, and the confined geometry of microchannels. 
By using shape-based analysis, we provide a new pathway for understanding and controlling droplet behavior in multiphase systems.

\section{\label{sec:level2}Methodology}

\subsection{Data base constructed from numerical study}

\subsubsection{Description of physical problem}

\begin{figure}[H]
	\centering

	\includegraphics[scale = 0.35]{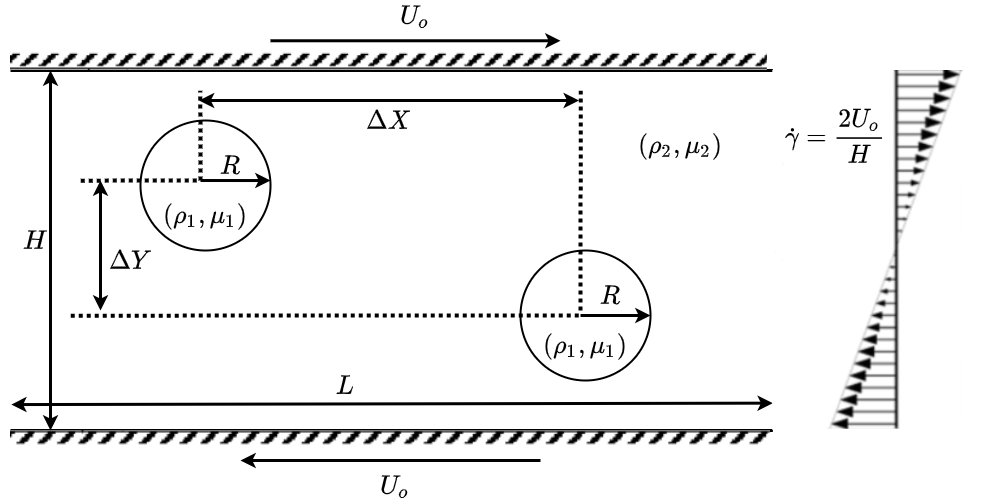}
		
	\caption{Schematic representation of a droplet pair with an initial radius of R in a confined shear flow. The droplets are located between two parallel plates that are a distance $H$ apart and move in opposite directions. $\Delta X$  and $\Delta Y$  are the horizontal and vertical distances between the centers of the droplets.}
	\label{fig:schematic}
\end{figure}

\begin{figure}[H]
          \centering

         \includegraphics[scale = 0.10]{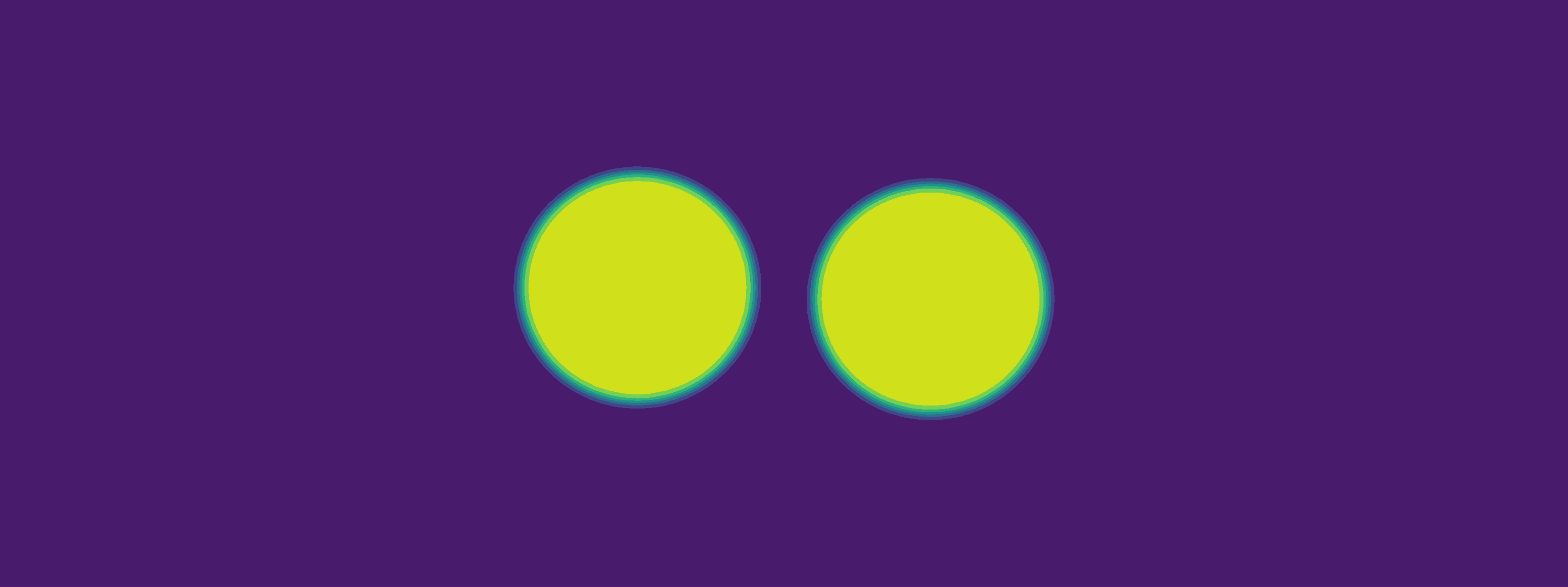}
		\includegraphics[scale = 0.10]{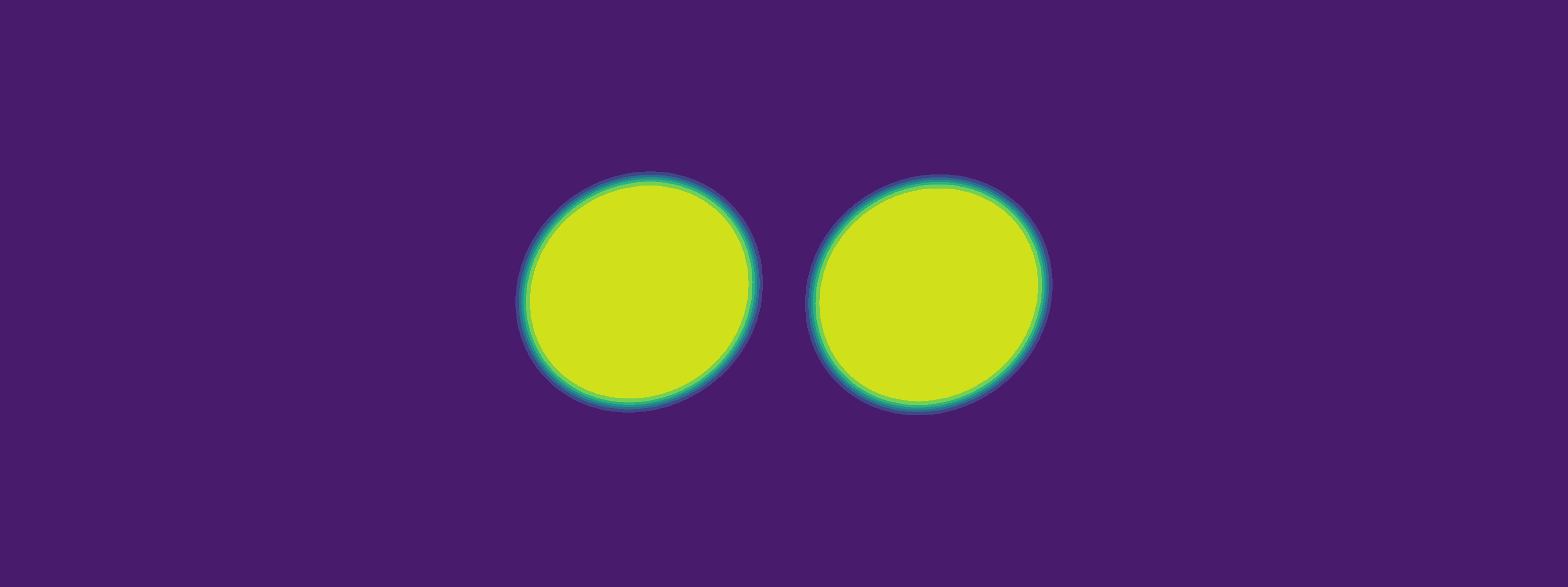}
		\includegraphics[scale = 0.10]{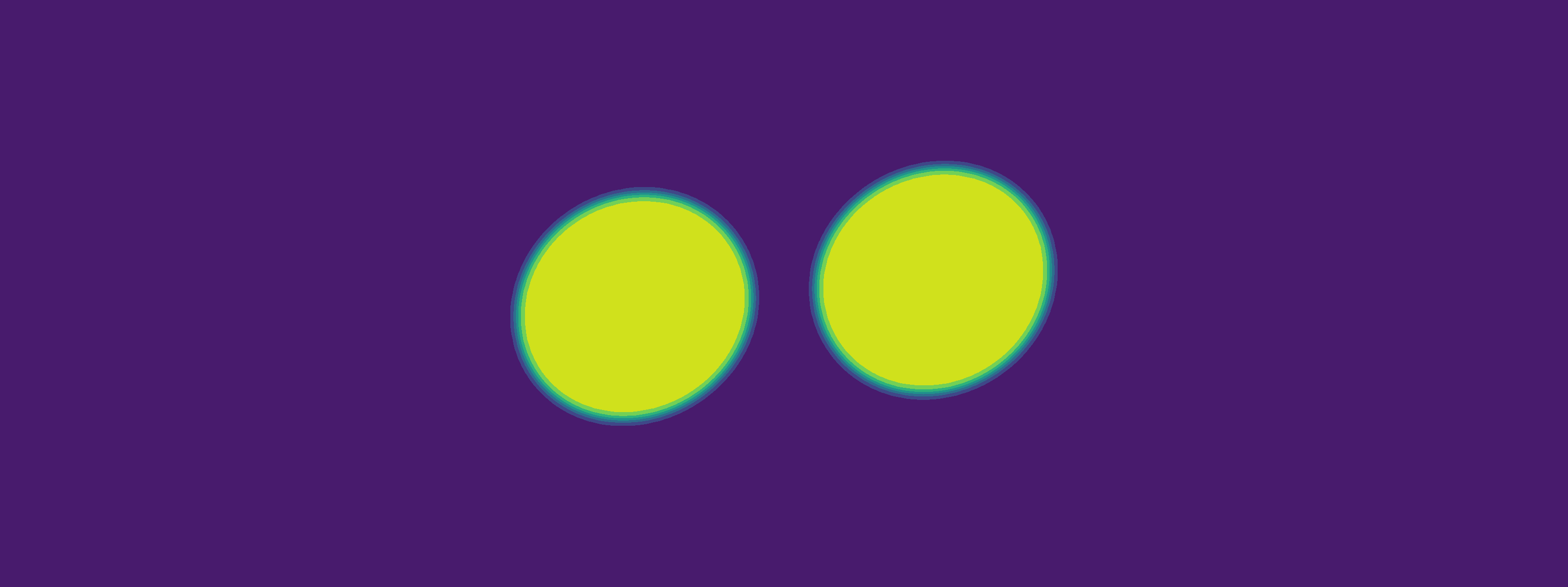}
		\includegraphics[scale = 0.10]{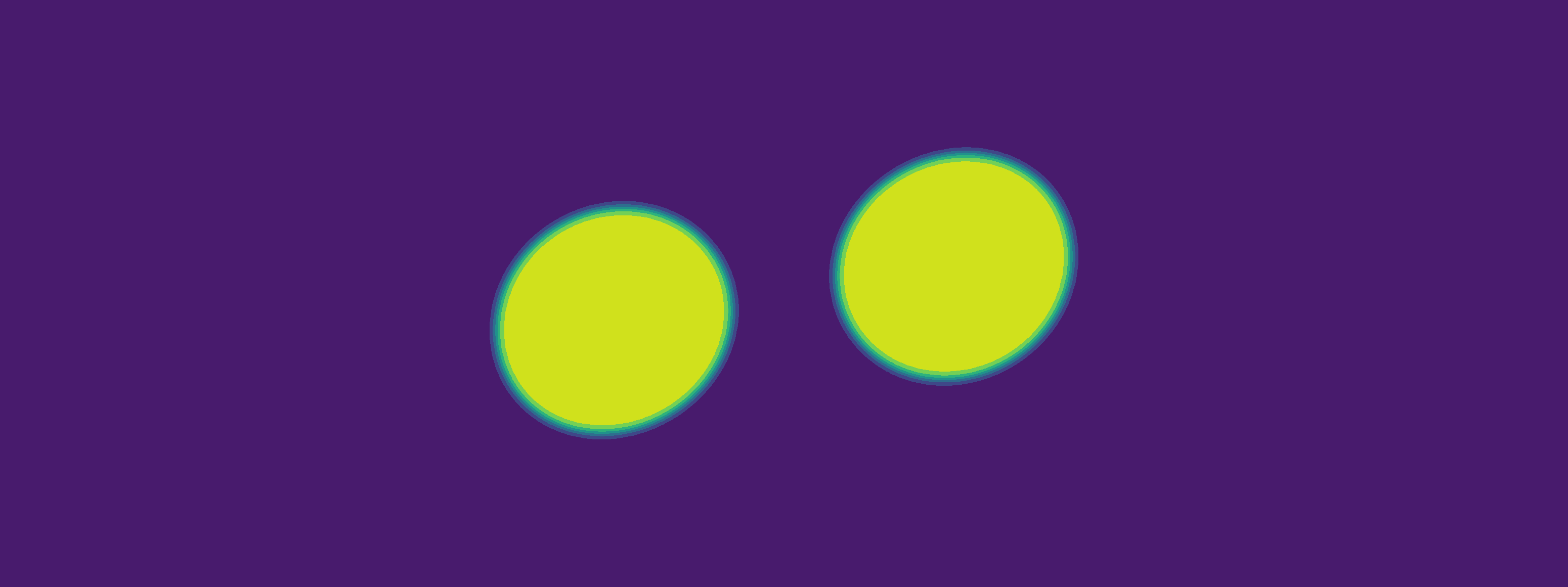}
		\includegraphics[scale = 0.10]{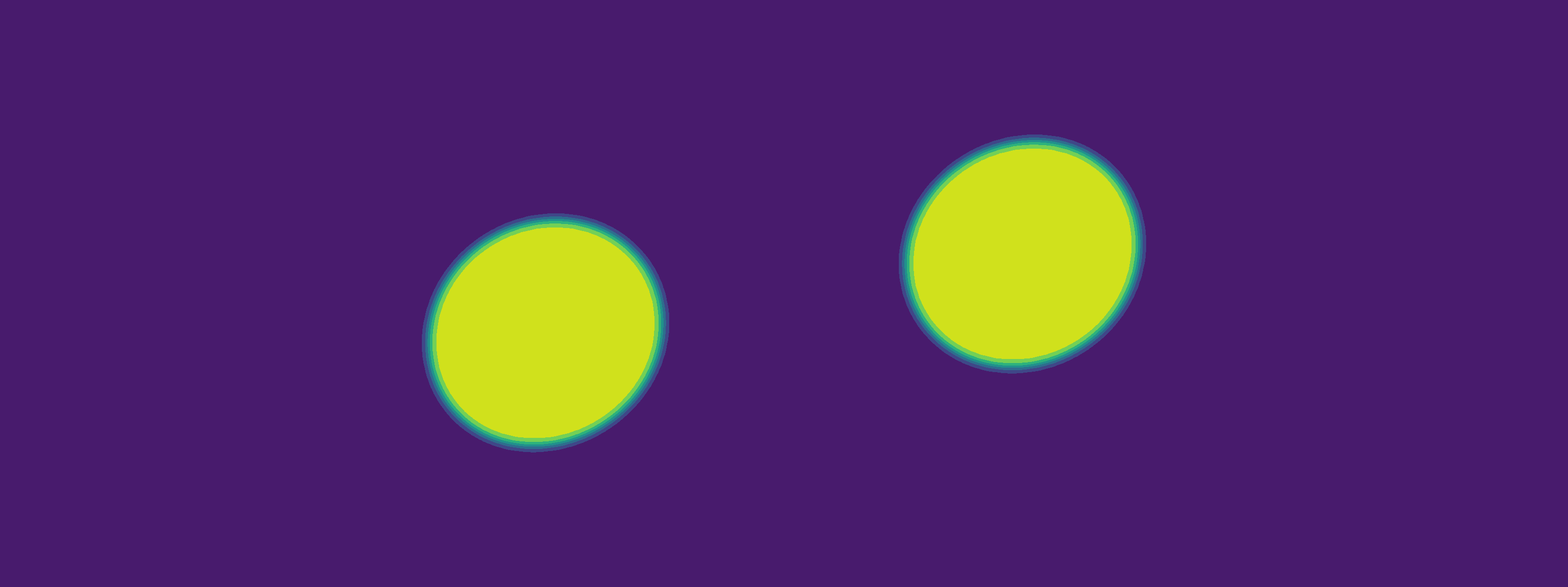}
		\\ (a)  Reverse-back \\ \vspace{0.3 cm}

        \includegraphics[scale = 0.10]{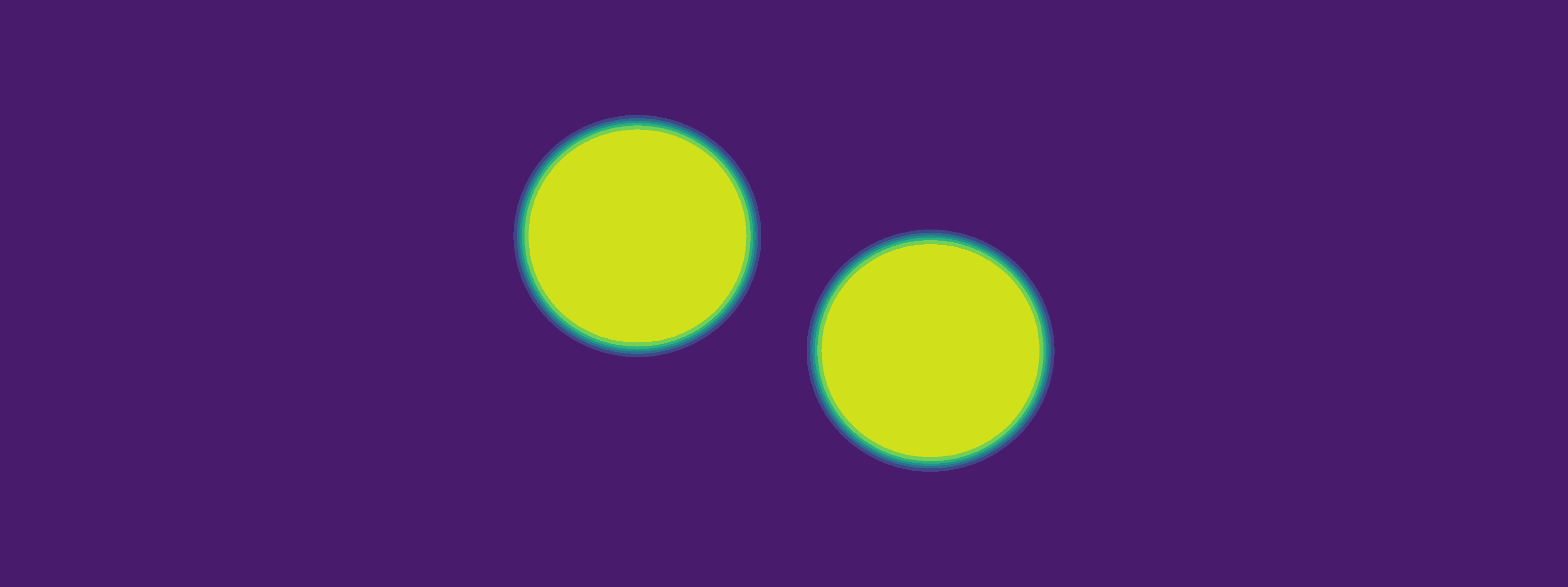}
            \includegraphics[scale = 0.10]{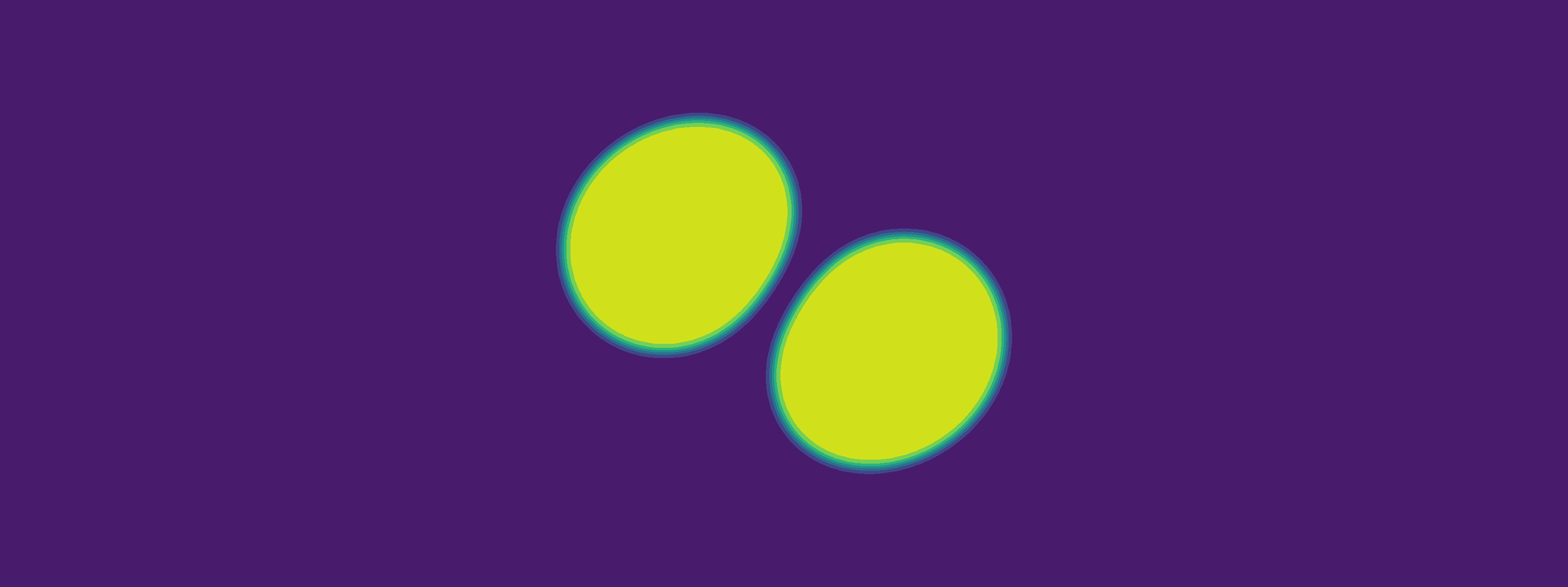}
		\includegraphics[scale = 0.10]{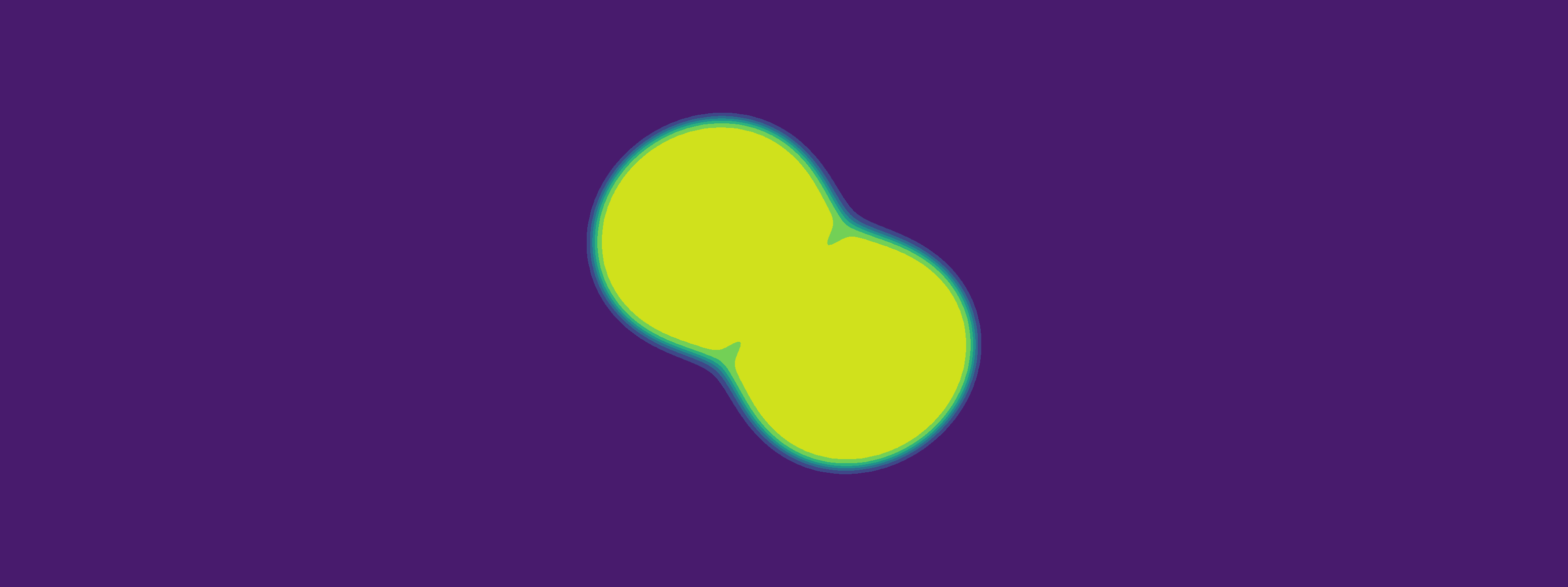}
		\includegraphics[scale = 0.10]{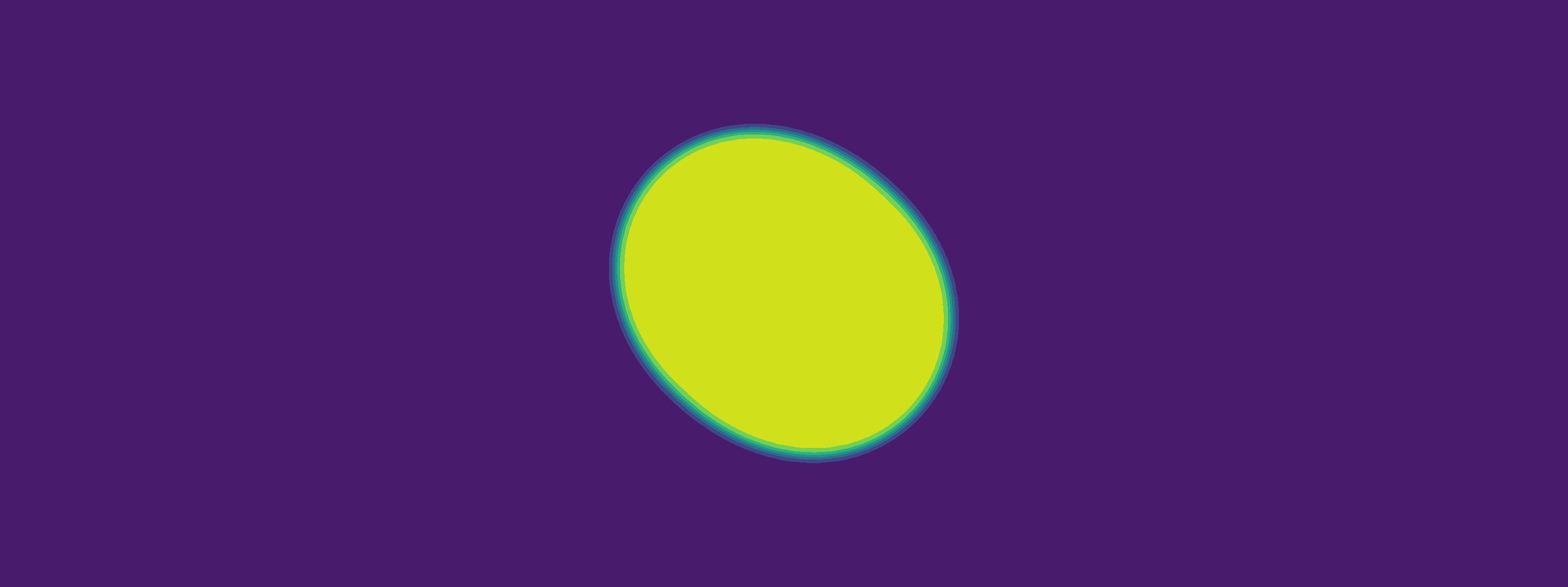}
		\includegraphics[scale = 0.10]{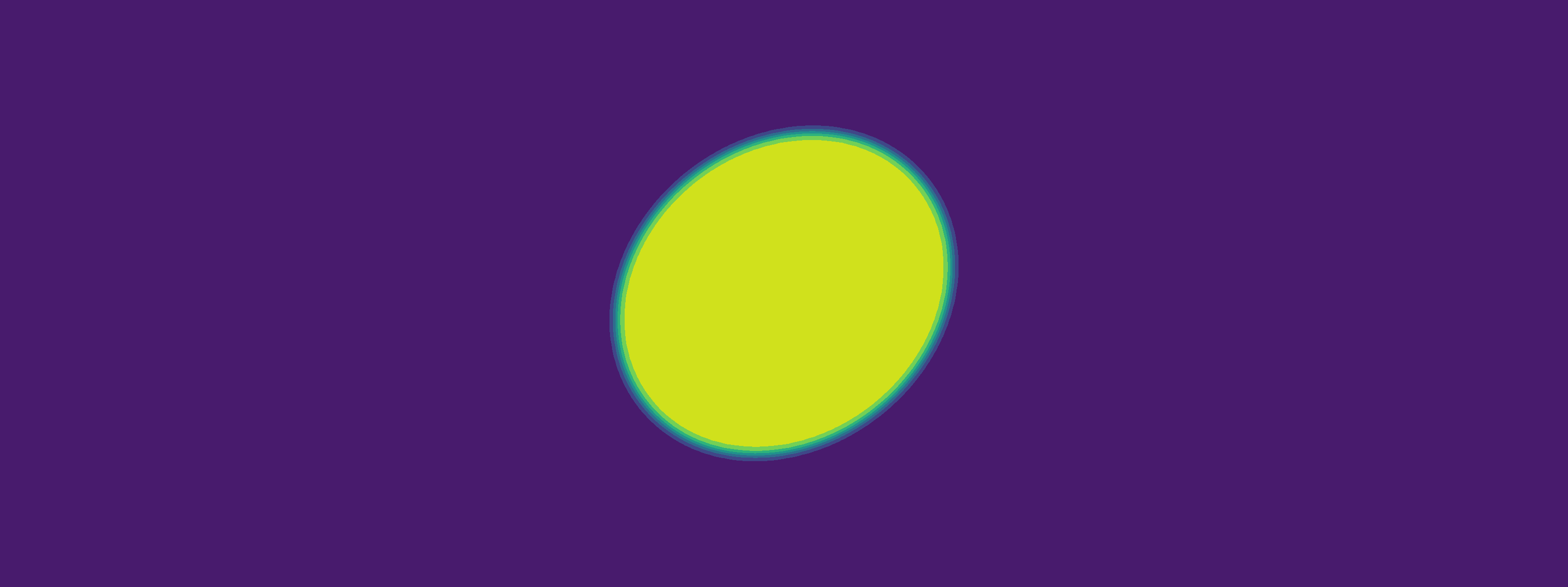}  
        \\ (b) Coalescence \\ \vspace{0.3 cm}

        \includegraphics[scale = 0.10]{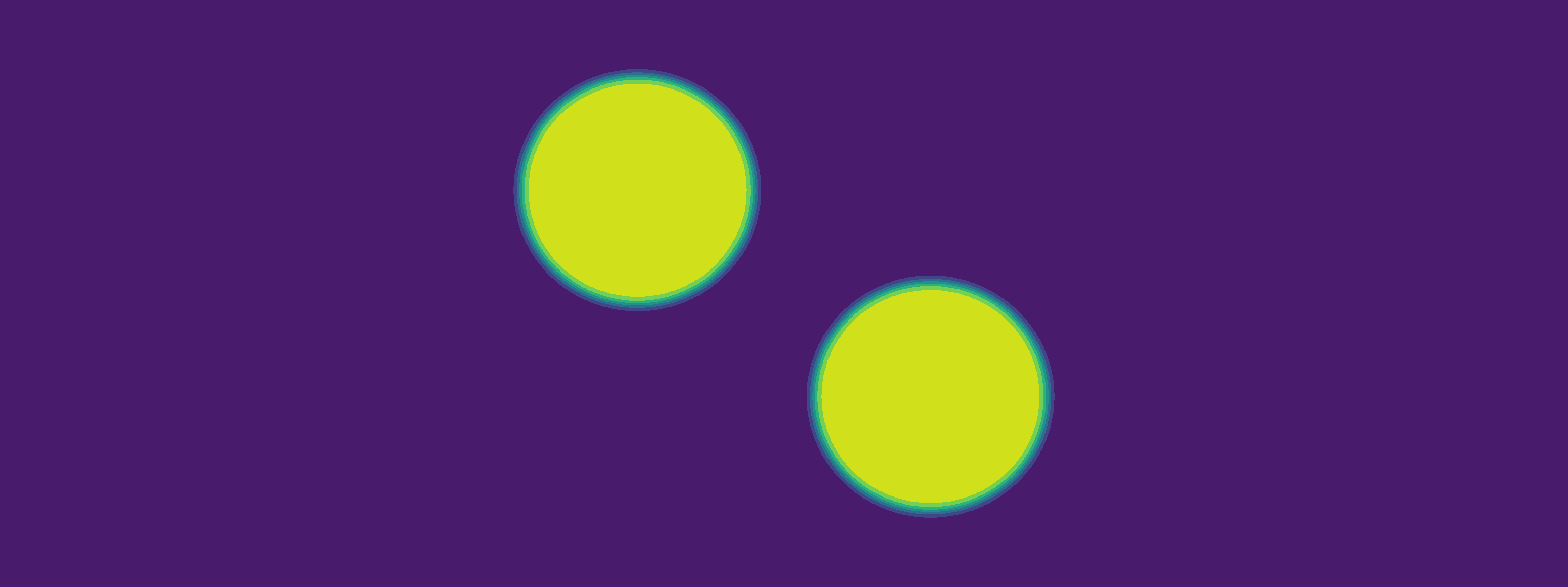}
            \includegraphics[scale = 0.10]{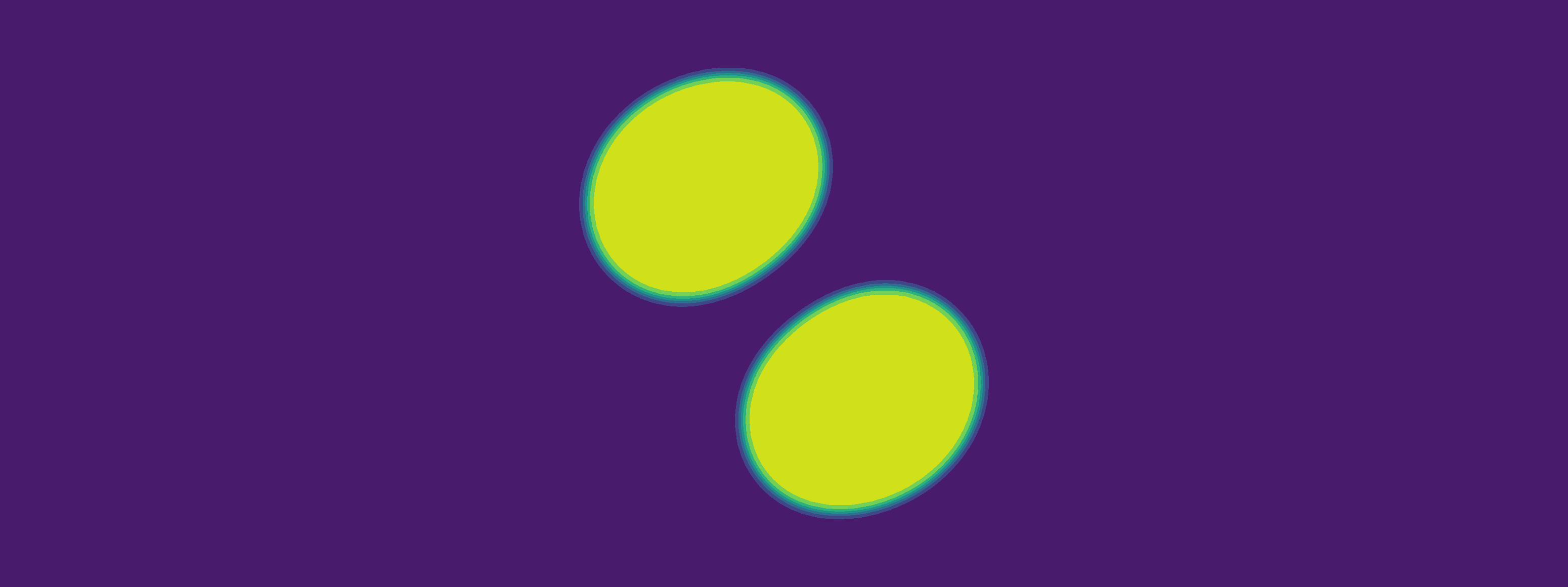}
		\includegraphics[scale = 0.10]{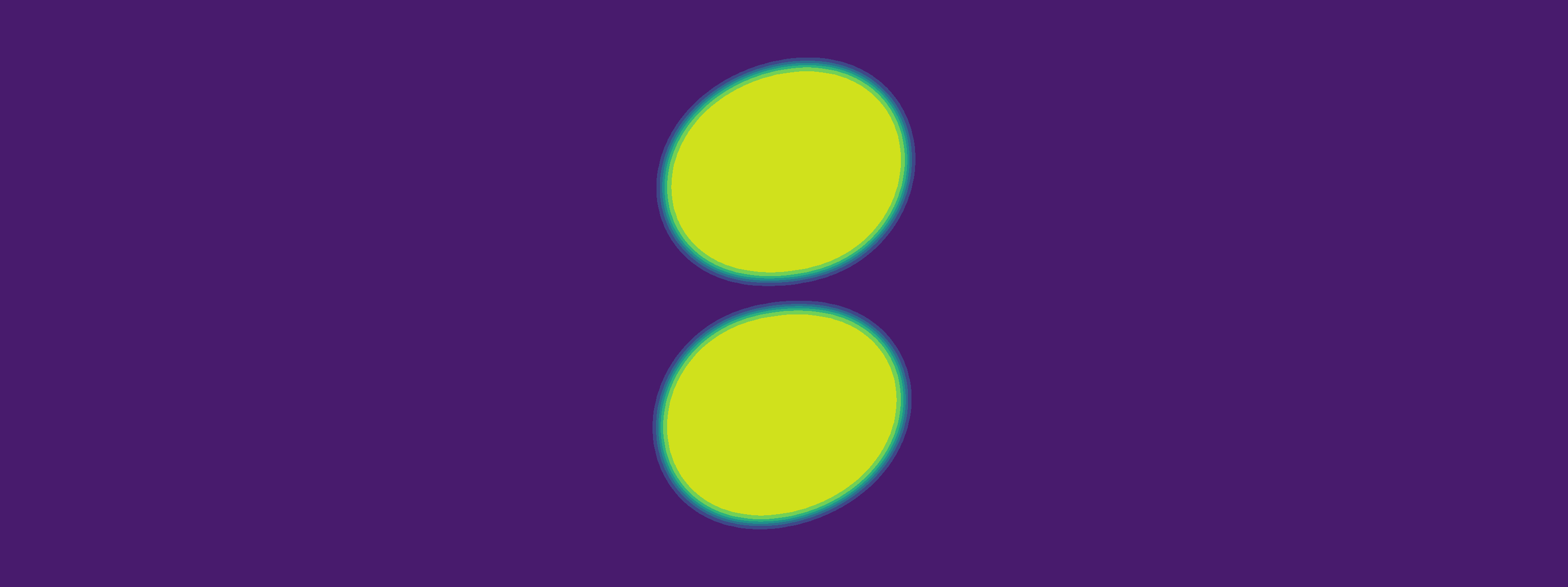}
		\includegraphics[scale = 0.10]{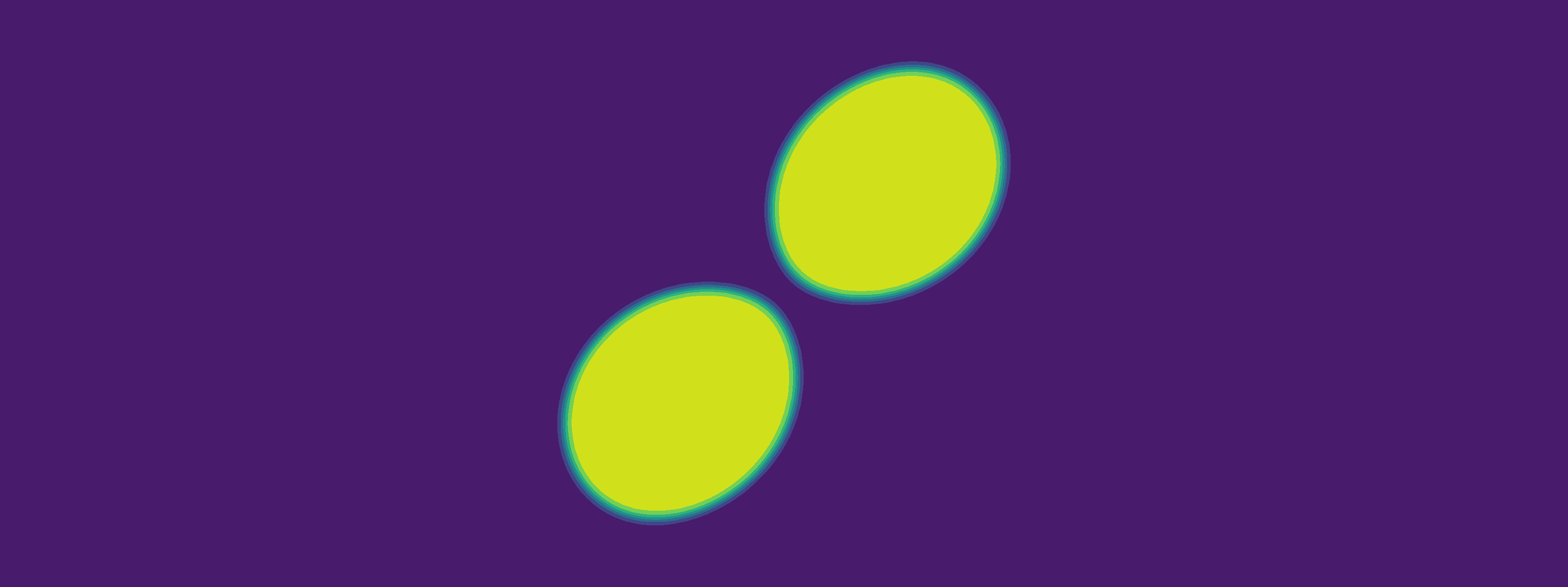}
		\includegraphics[scale = 0.10]{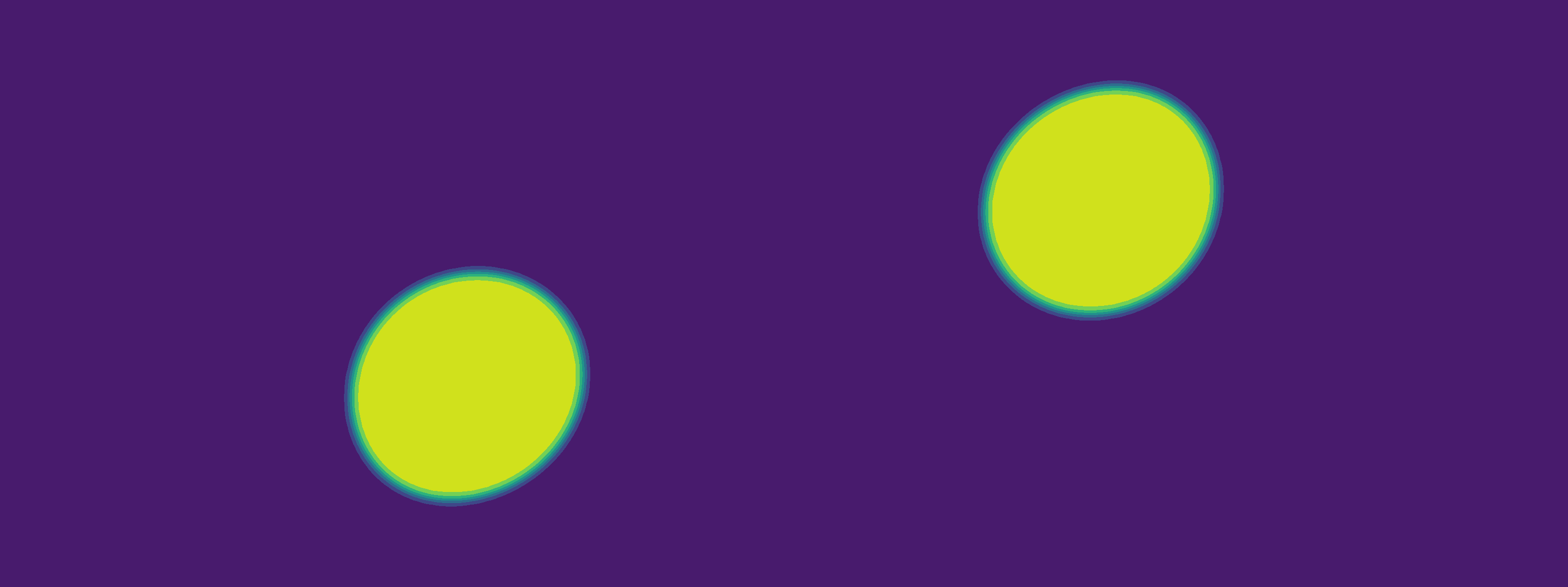}  
       \\ (c) Pass-over \\ \vspace{0.3 cm}
       
 	\caption{Demonstration of three different cases with three different outcomes: reverse-back, coalescence, and pass-over. The simulation parameters considered  are $Ca=0.01$, $Re=1.0$, $\rho_{12}=1$, $\mu_{12}=1$, $2R/H=0.39$, with (a) $(\Delta Y/2R)_{in}=$ 0.05 (reverse-back), (b) $(\Delta Y/2R)_{in}=$0.50 (coalescence), and (c) $(\Delta Y/2R)_{in}=$0.90 (pass-over).} 
	\label{fig:demo_three_different_cases}
\end{figure}

Our problem case involve simple droplet pair collision in a confined shear flow. A schematic two-dimensional representation for the collision of droplet pairs in a confined shear flow is demonstrated in Figure~\ref{fig:schematic}. Two spherical droplets, both of radius R, are initially separated by a horizontal spacing of $\Delta X$ and a vertical distance of $\Delta Y$ as measured between their centers. The simple shear flow is generated by the motion of the top and bottom walls with a shear rate of $\dot{\gamma}=2U_o/H$, where $U_o$ is the horizontal speed of the planes as indicated in Figure~\ref{fig:schematic}, and $H$ is the vertical spacing between them. Other parameters that define the physical problem are the densities of the droplet ($\rho_1$) and matrix fluid ($\rho_2$), the dynamic viscosities of the droplet ($\mu_1$) and matrix fluid ($\mu_2$), and the surface tension ($\sigma$) between the droplet and matrix fluid. The dimensionless parameters that play critical roles in characterizing the collision outcome of the droplet pairs are the Reynolds number ($Re =  \frac{\rho_2 \dot{\gamma }R^2}{\mu_2 }$, which quantifies the relative importance between the inertia force and the viscous force), the Capillary number ($Ca= \frac{\mu_2 \dot{\gamma }R}{\sigma}$, which shows the ratio of the viscous force to the surface tension force), density ratio ($\rho_{12}=\frac{\rho_1}{\rho_2}$), viscosity ratio ($\mu_{12}=\frac{\mu_1}{\mu_2}$), dimensionless vertical initial offset $((\Delta Y/2R)_{in})$, and the confinement $(2R/H)$. Also for a brief overview, the evolution of three distinct outcomes (reverse-back, coalescence, and pass-over) from droplet pair collisions, influenced by different parameters, is demonstrated through simulated snapshots presented in Figure~\ref{fig:demo_three_different_cases}.

\subsubsection{Numerical simulations}
To generate the required data numerically for the collision of pair droplets in confined shear flows, we have adopted a lattice Boltzmann method (LBM)~\cite{lee2010lattice}, based on the Cahn-Hilliard diffuse interface theory for binary immiscible fluids. In the LBM for binary fluid, one follows the evolution of two discrete density distributions, $g_{\alpha}$ and $h_{\alpha}$, to model hydrodynamics and the evolution of phase-field, respectively.
We will be establishing a brief overview of the method here; for more details on the theory and method, readers are referred to Refs.~\cite{lee2010lattice, mohamad2011lattice,bhatnagar1954model,he1998discrete,lee2009effects,lee2014physical}. A scalar order parameter $C$ (i.e., composition) is introduced, which satisfies the following convective Cahn-Hilliard equation~\cite{lee2010lattice}:
\begin{equation}
    \frac{\partial C}{\partial t}+\mathbf{u} \cdot \nabla C =M {\nabla}^2 \mu_c,
    \label{CH-2}
\end{equation}
which involves mobility parameter $M$ and chemical potential $\mu_c$. Among the two distribution functions $g_{\alpha}$ and $h_{\alpha}$, the former is used for the calculation of pressure and momentum of the two-component mixture, and the latter is used as a phase-field function for the transport of the composition. The discrete Boltzmann equations for pressure evolution and momentum equations and advective
phase-field equation are, respectively~\cite{lee2010lattice}:

\begin{equation}
\frac{\partial g_{\alpha}}{\partial t} + \mathbf{e}_{\alpha} \cdot \nabla g_{\alpha} = - \frac{1}{\lambda}(g_{\alpha} - g_{\alpha}^{eq}) 
+ (\mathbf{e}_{\alpha} - \mathbf{u}) \cdot [\nabla \rho c_s^2 (\Gamma_{\alpha}-\Gamma_{\alpha}(0))+ \mu_c \nabla C \Gamma_{\alpha} ]
\end{equation}
\begin{equation}
\frac{\partial h_{\alpha}}{\partial t} + \mathbf{e}_{\alpha} \cdot \nabla h_{\alpha} =  -\frac{1}{\lambda} (h_{\alpha}-h_{\alpha}^{eq}) + M\nabla^2 \mu_c \Gamma_{\alpha} + (\mathbf{e}_{\alpha}-\mathbf{u}) \cdot [\nabla C - \frac{C}{\rho c_s^2} (\nabla p - \mu_c \nabla C)]\Gamma_{\alpha},
\end{equation}
where the equilibrium distribution functions are defined as:
\begin{equation}
g_{\alpha}^{eq} = t_{\alpha} \left [ p + \rho c_s^2 \left ( \frac{\mathbf{e}_{\alpha}\cdot \mathbf{u}}{c_s^2} + \frac{(\mathbf{e}_{\alpha}\cdot \mathbf{u})^2}{2c_s^4} - 
\frac{\mathbf{u}\cdot \mathbf{u}}{2c_s^2} \right )
\right ]
\end{equation}
\begin{equation}
h_{\alpha}^{eq} = t_{\alpha} C \left [ 1+ \frac{\mathbf{e}_{\alpha}\cdot \mathbf{u}}{c_s^2} + \frac{(\mathbf{e}_{\alpha}\cdot \mathbf{u})^2}{2c_s^4} - 
\frac{\mathbf{u}\cdot \mathbf{u}}{2c_s^2} 
\right ].
\end{equation}
In the above equations, $\mathbf{e}_{\alpha}$ is the microscopic particle velocity in the $\alpha$ direction, $\mathbf{u}$ is the volume-averaged velocity, $c_s$ is the lattice speed of sound, $\rho$ is the mixture density, $p$ is the dynamic pressure, $\lambda$ is the relaxation time, $t_{\alpha}$ is the weighting
factor, and $\Gamma_{\alpha}$ is defined as $\Gamma_{\alpha}=  \Gamma_{\alpha}(\textbf{u})=h_{\alpha}^{eq}/C$. 
The dimensionless relaxation time $\tau=\lambda/\delta t$ is related to the kinematic viscosity by $\nu=\tau c_s^2 \delta t$.
The density can be related to the composition using the following linear function~\cite{lee2010lattice}:
\begin{equation}
    \rho = C \rho_1 + (1-C)\rho_2,
\end{equation}
where $\rho_1$ and $\rho_2$ are the bulk densities of two fluids. The chemical potential, $\mu_{c}$ is defined as the derivative of the free energy with respect to the order parameter, and the free energy is given by~\cite{lee2014physical}:
\begin{equation}
    E(C) = E_0(C) + \frac{\kappa}{2} \left | \nabla C \right |^2,
\end{equation}
where $E(C)$ is the total free energy, $E_0(C)$ is the free energy density of the binary solution, $\nabla C$ is the composition gradient, and $\kappa$ is the gradient parameter related to the surface tension of the interface between two phases. The free energy density is taken as $E_0(C) = \beta C^2(C-1)^2$~\cite{lee2010lattice} with $\beta$ being a constant. The equilibrium profile and surface tension of the interface in equilibrium can be determined by minimizing the mixing energy. The plane interface profile at equilibrium is then evaluated as~\cite{lee2010lattice}:
\begin{equation}
    C(z) = \frac{1}{2}+\frac{1}{2}\text{tanh}(\frac{2z}{\xi}),
\end{equation}
where $z$ denotes the coordinate normal to the interface plane, and $\xi$ is the interface thickness. Once the surface tension and interface thickness are chosen, $\beta$ and $\kappa$ are specified as $\beta=12\sigma/\xi$ and $\kappa=\beta \xi^2/8$.
Finally, the macroscopic variables such as composition, momentum, and dynamic pressure can be obtained by taking the moments of $h_{\alpha}$ and $g_{\alpha}$:
\begin{equation}
    C = \sum_{\alpha}h_\alpha ,
\end{equation}
\begin{equation}
    \rho \mathbf{u} = \frac{1}{c_s^2}\sum_{\alpha} \mathbf{e}_{\alpha}g_{\alpha},
\end{equation}
\begin{equation}
    p = \sum_{\alpha} g_{\alpha}.
\end{equation}
For comprehensive discretization of
Eqs.(2) and (3) and boundary conditions, readers are referred to Ref.~\cite{lee2010lattice}. 

This numerical method has been extensively validated in our previous works, where it has been successfully applied to study various interactions of droplets, including their pair-wise collisions in confined shear flows~\cite{al2024hydrodynamic, PhysRevFluids.7.123603, farokhirad2013effects}, as well as their interactions with solid substrates~\cite{farokhirad2023spreading, farokhirad2014pof, farokhirad2015coalescence, farokhirad2017coalescence, farokhirad2017multiphase}. 
Building on this validated framework, the current work leverages the LBM to generate high-fidelity simulation data, which serves as the input for developing ML models aimed at predicting the collision outcomes of droplet pairs in confined shear flows.

\subsection{Convolutional neural network modeling}

\subsubsection{Data collection and preprocessing}
The vertical initial offset and channel confinement, two pivotal geometric parameters, exert a crucial influence on producing different orientations and deformations of droplets, ultimately leading to variations in their outcomes. We strategically selected 22 cases of vertical initial offset, covering a range from 0.05 to 1.10 with increments of 0.05 (denoted as $\frac{\Delta Y}{2R}$). For each case, we considered 10 unique channel confinement values, ranging from 0.30 to 0.39 with an increment of 0.01 (expressed as $\frac{2R}{H}$), resulting in a total of 220 cases for training. From each cases, we extracted random snapshots in close proximity to the collision event, totaling 3010 snapshots. The dataset comprises 680 snapshots from the reverse-back collision case, 1430 snapshots from the coalescence case, and 900 snapshots from the pass-over case. These images serve as the input data for training the Convolutional Neural Network (CNN) model and each snapshot is labeled accordingly. The CNN model is thereby trained to recognize and learn critical features associated with the shape and positioning of droplets during collisions, enabling it to distinguish between the different collision outcomes based on these key characteristics.

\begin{figure}[H]
	\centering

		\includegraphics[scale = 0.045]{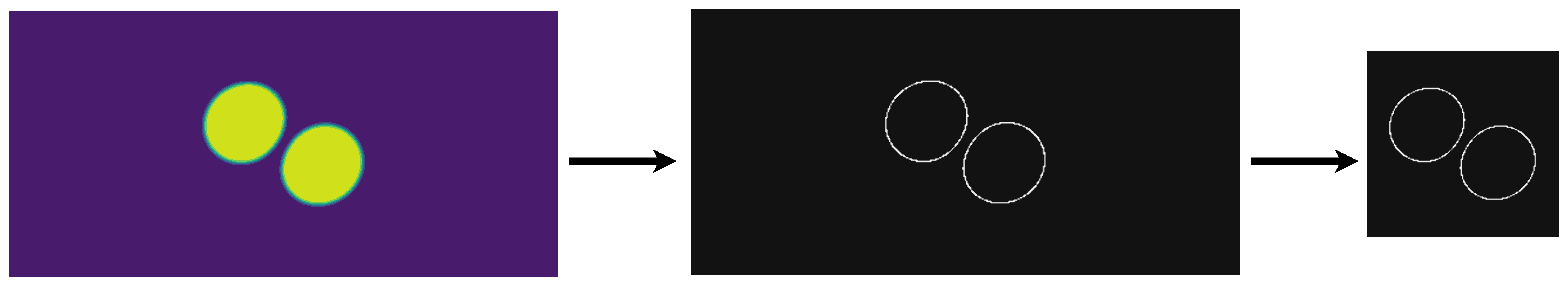} \\ \vspace{0.3cm}
		
	\caption{Cropping process for reducing dimension.}
	\label{fig:cropping}
\end{figure}

\begin{figure}[H]
          \centering

         \includegraphics[scale = 0.18]{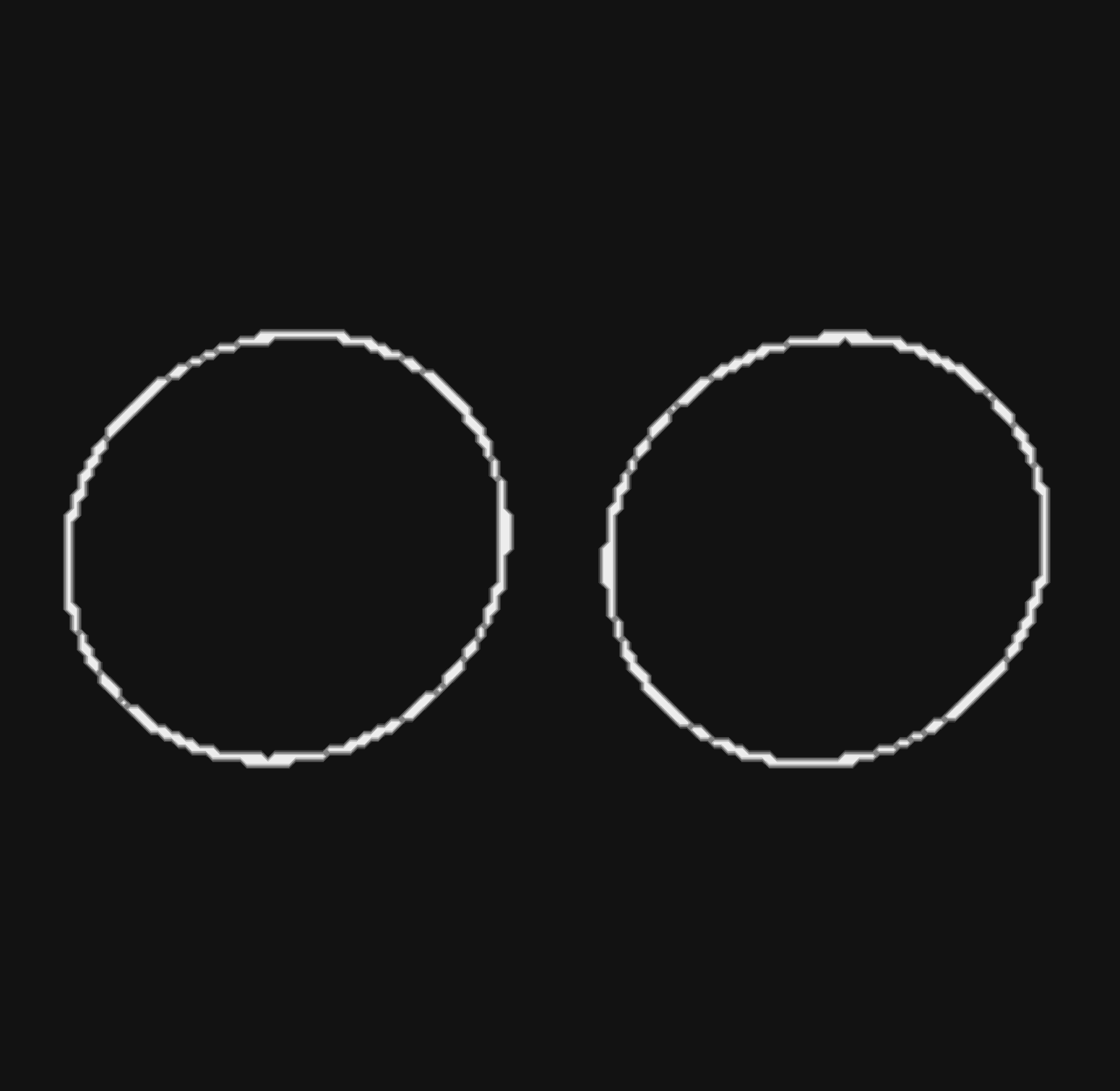}
		\includegraphics[scale = 0.18]{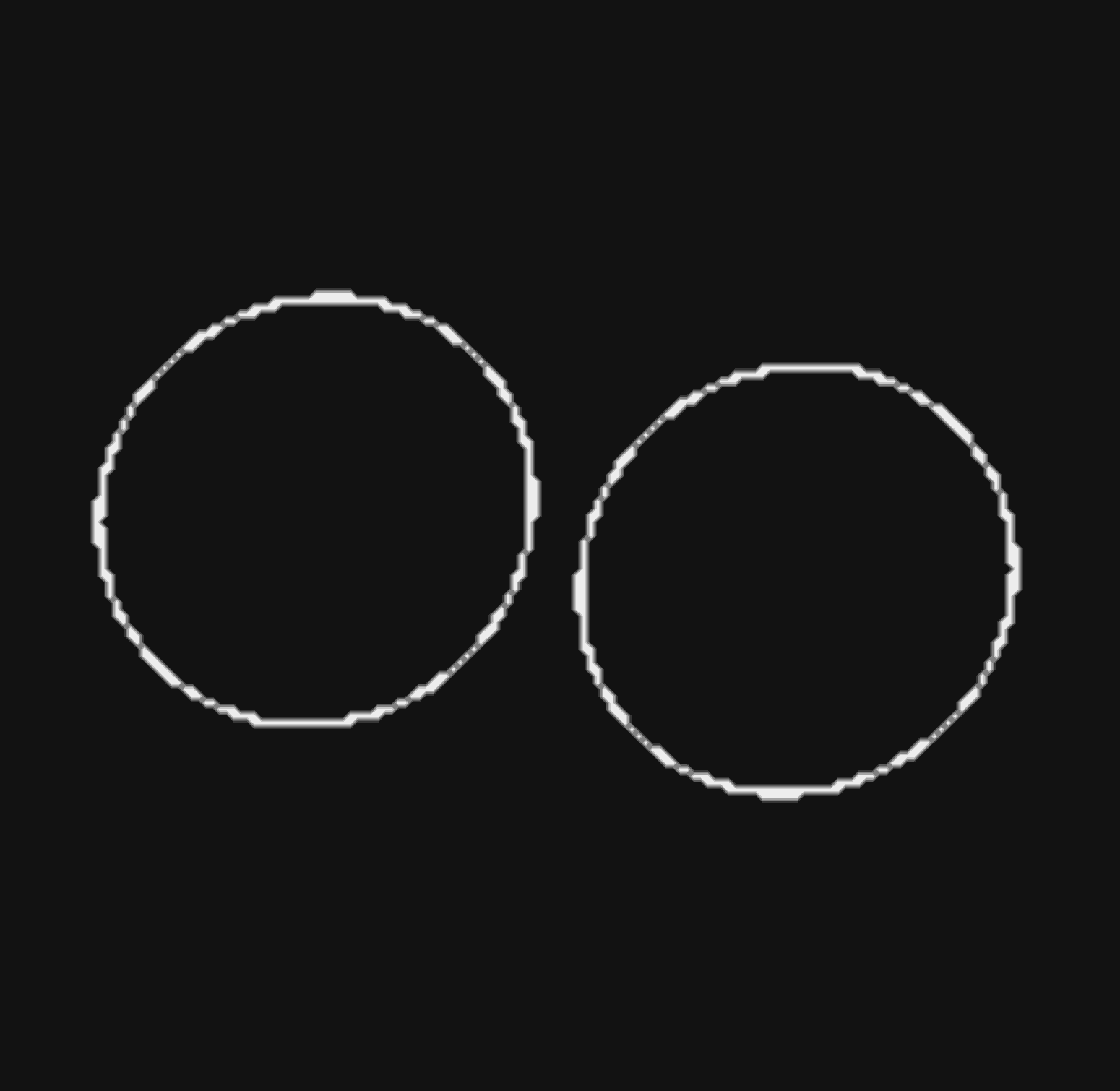}
		\includegraphics[scale = 0.18]{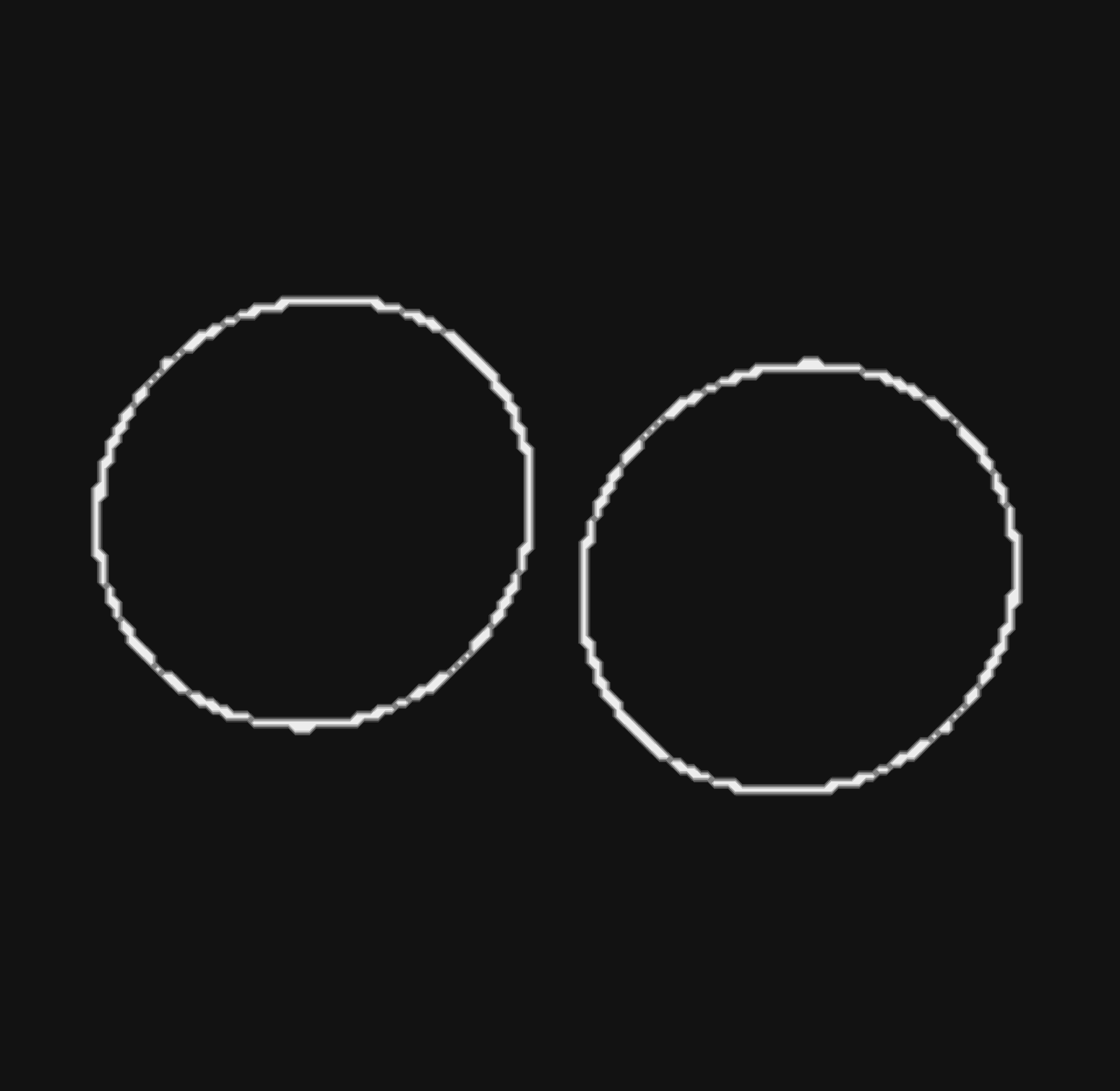}
		\includegraphics[scale = 0.18]{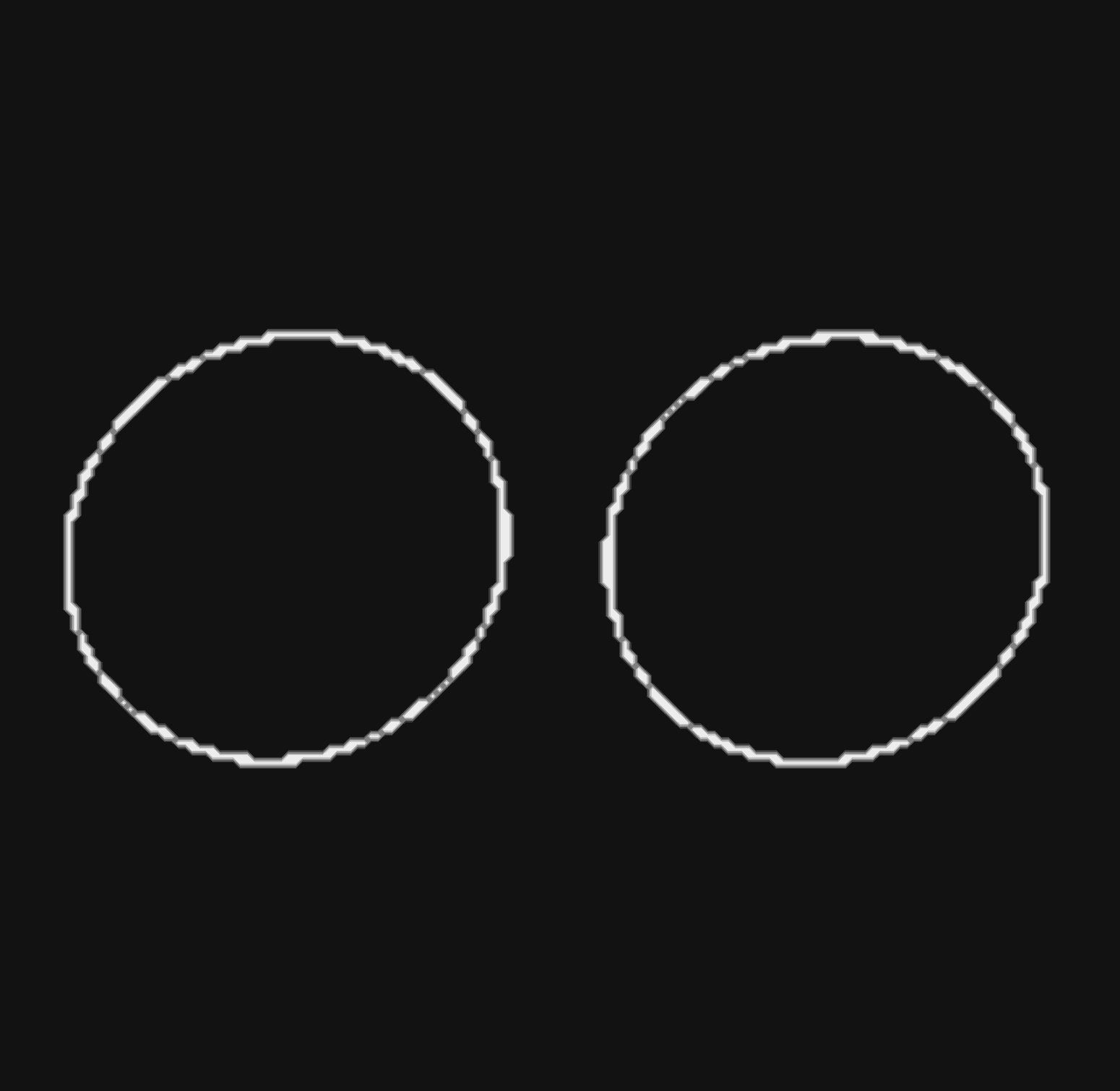}
		\includegraphics[scale = 0.18]{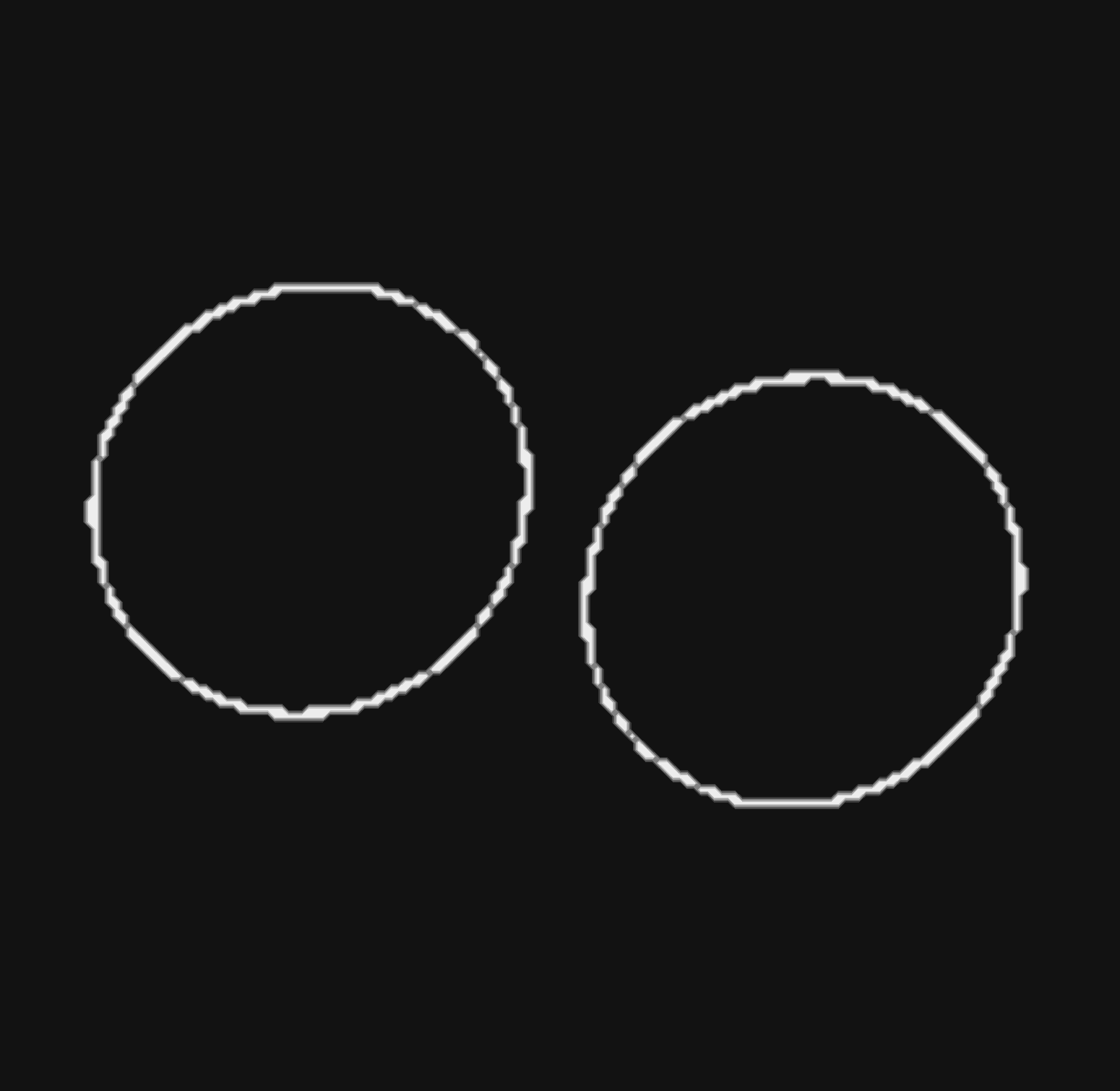}
		\\ (a)  Reverse-back \\ \vspace{0.3 cm}

        \includegraphics[scale = 0.18]{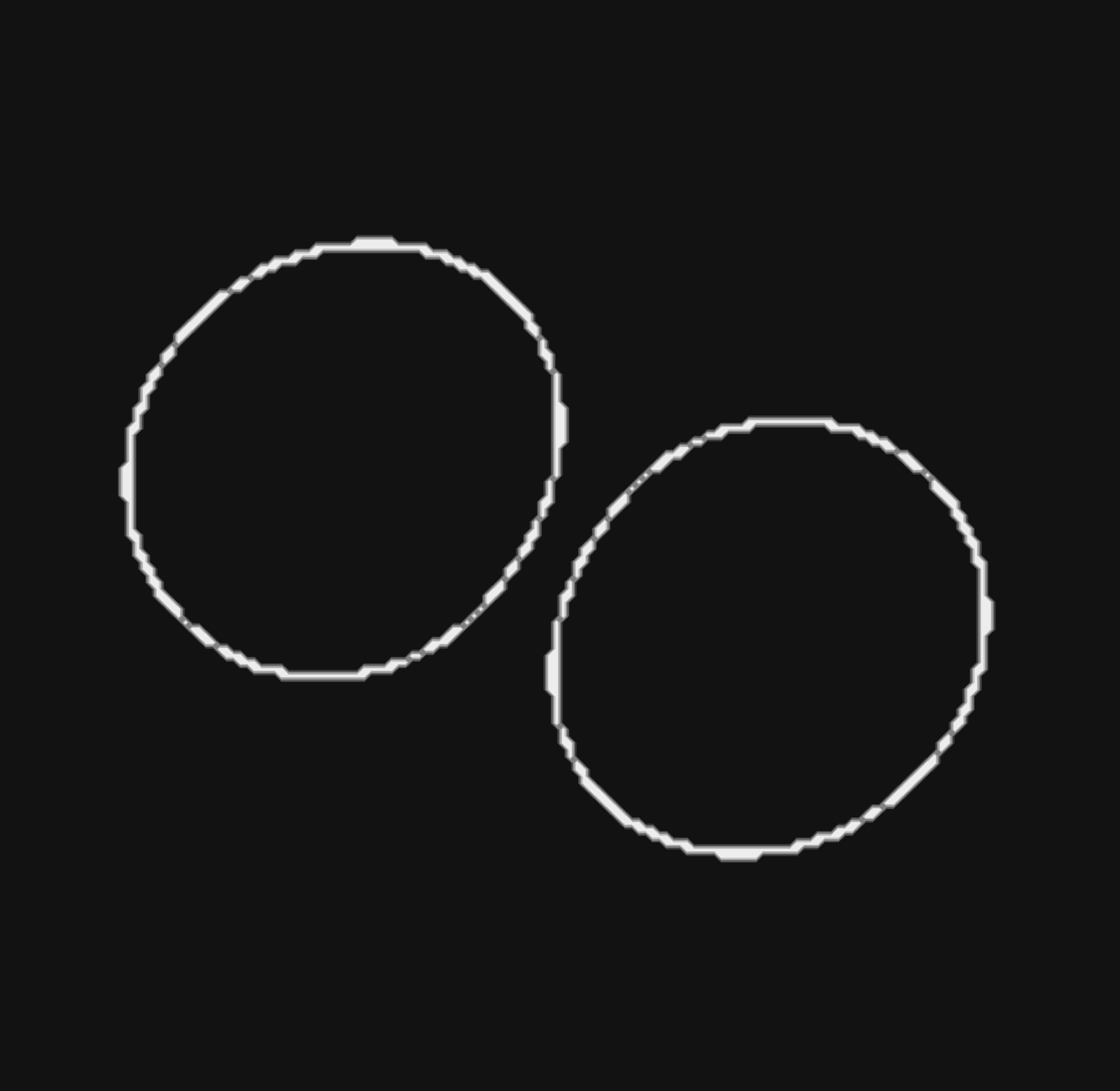}
            \includegraphics[scale = 0.18]{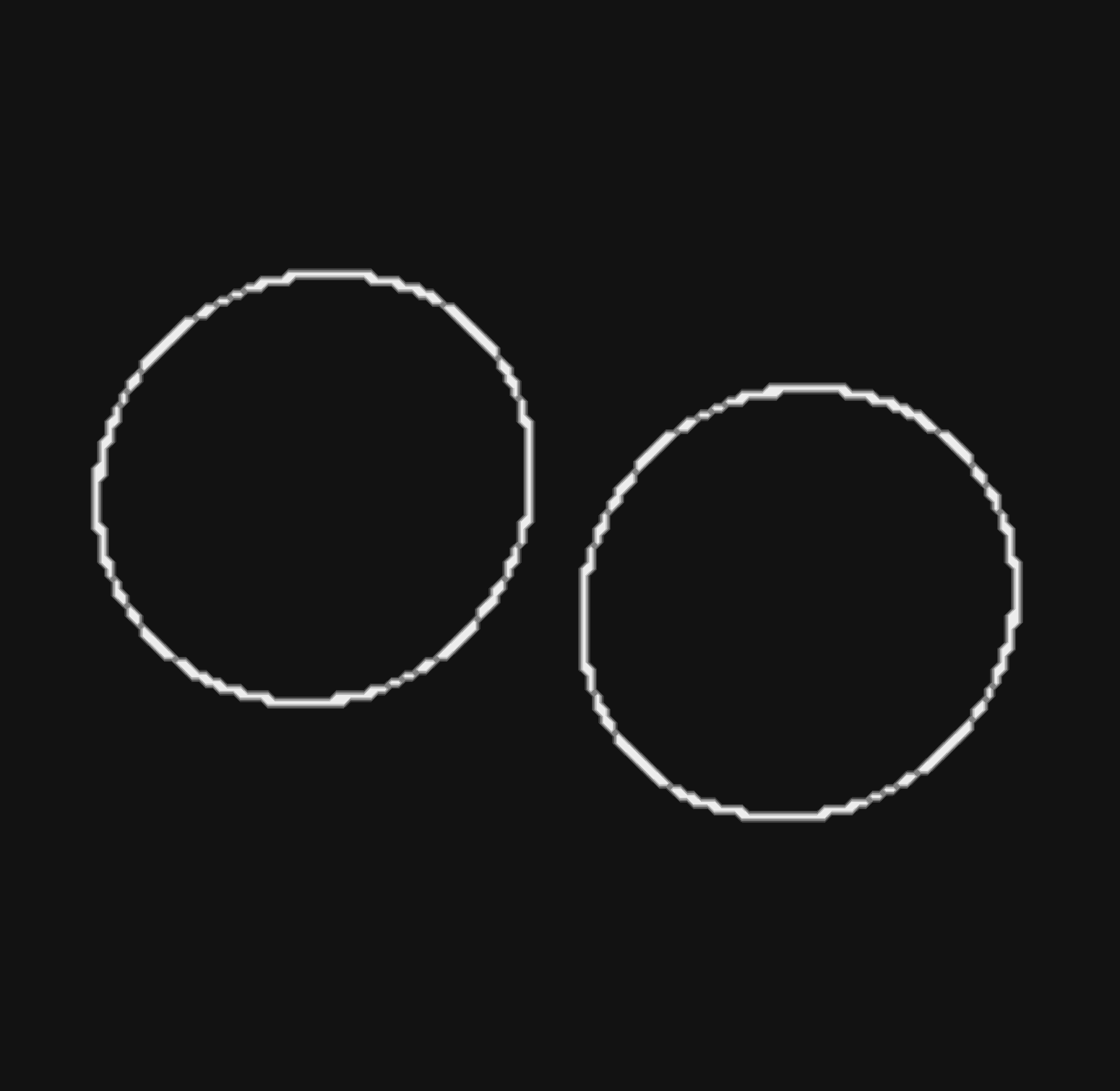}
		\includegraphics[scale = 0.18]{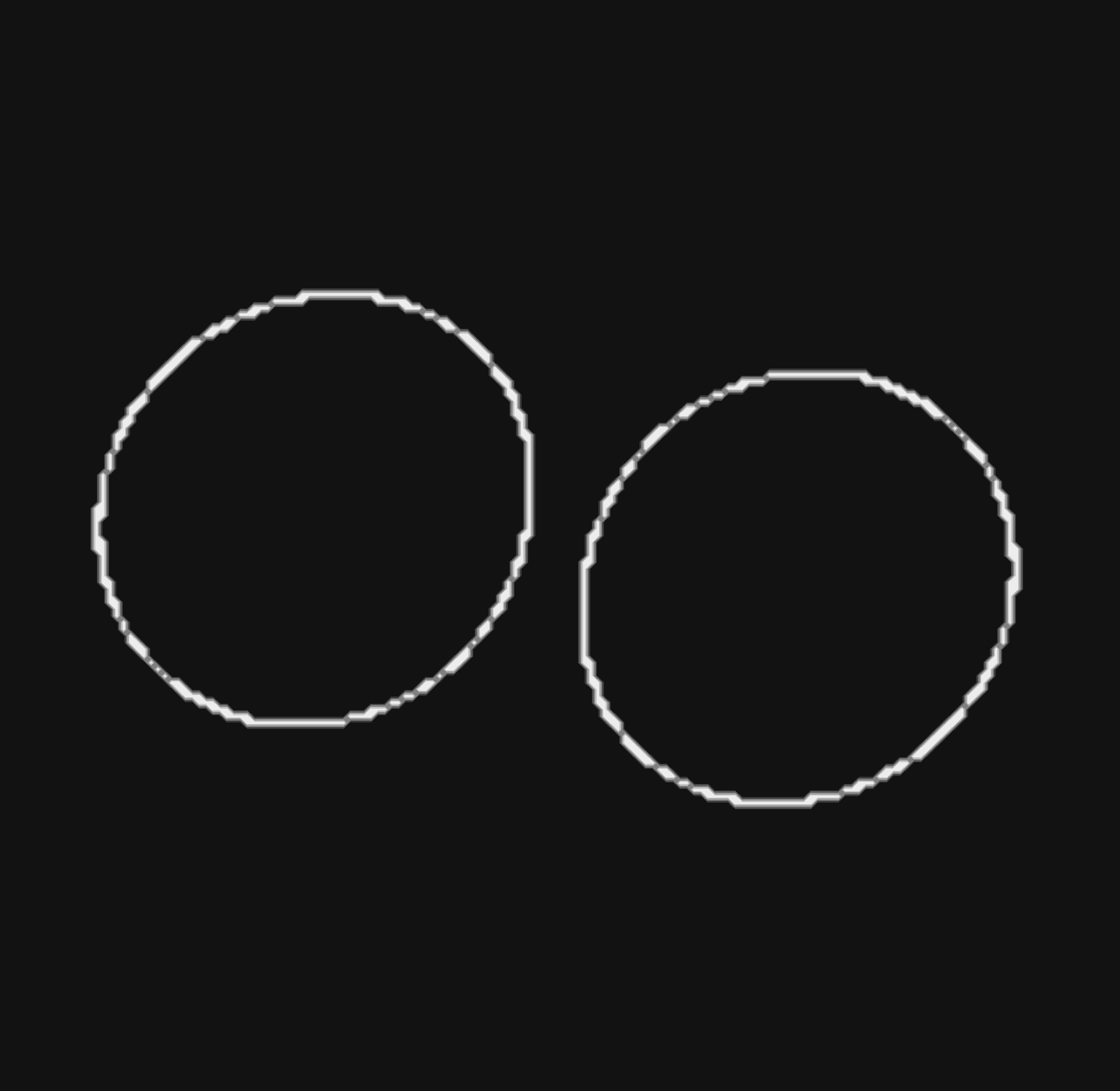}
		\includegraphics[scale = 0.18]{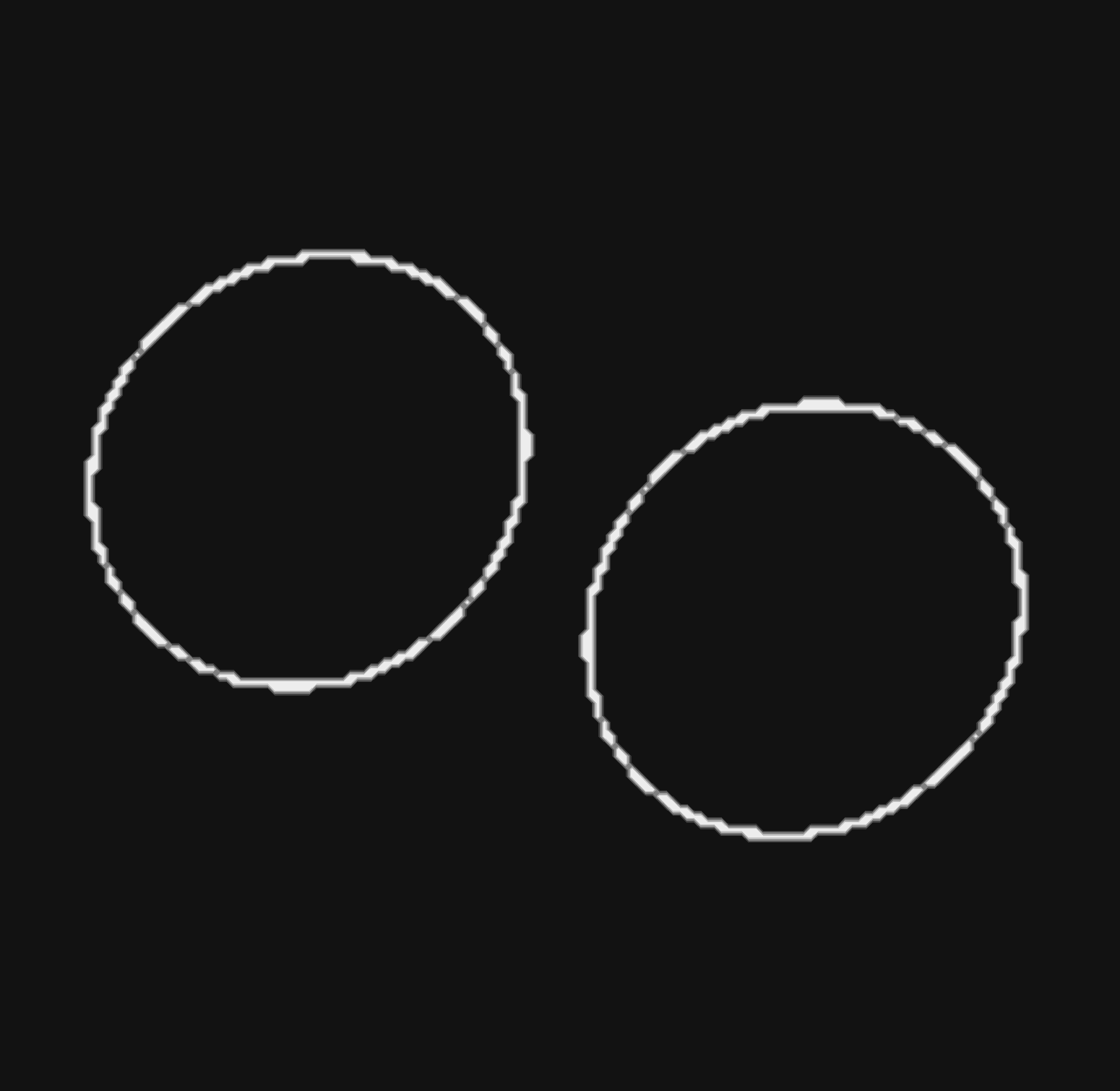}
		\includegraphics[scale = 0.18]{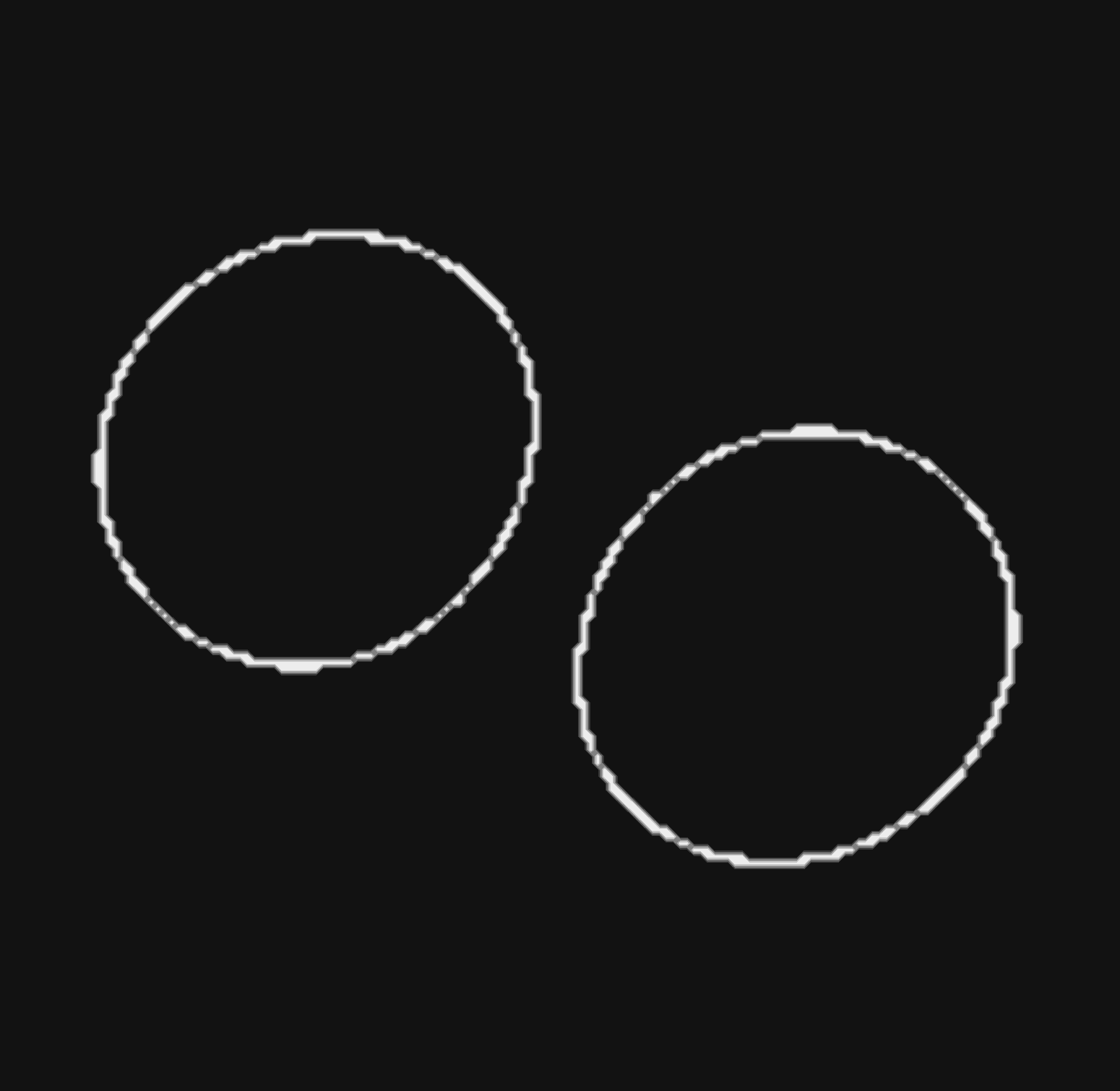}  
        \\ (b) Coalescence \\ \vspace{0.3 cm}

        \includegraphics[scale = 0.18]{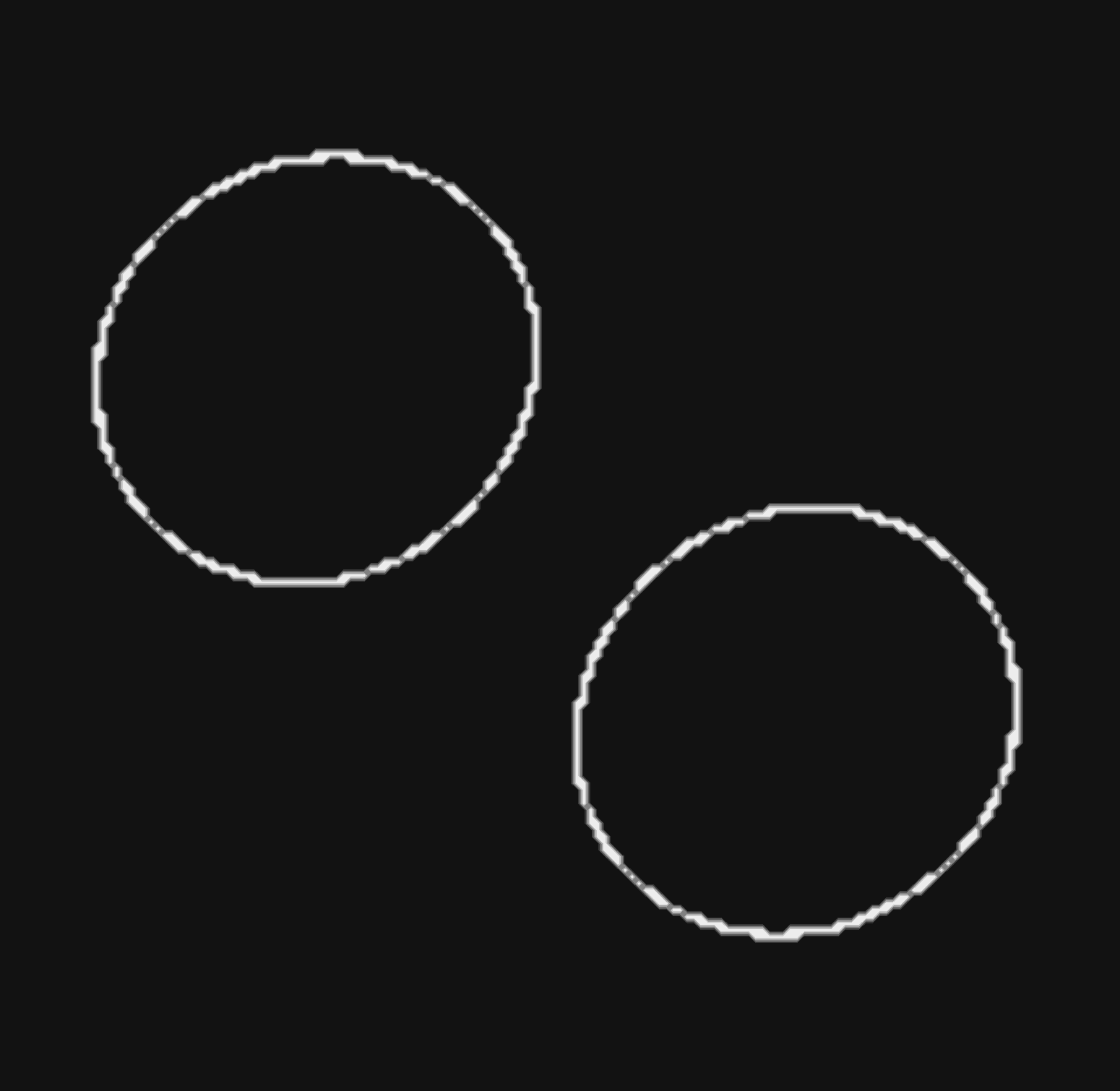}
            \includegraphics[scale = 0.18]{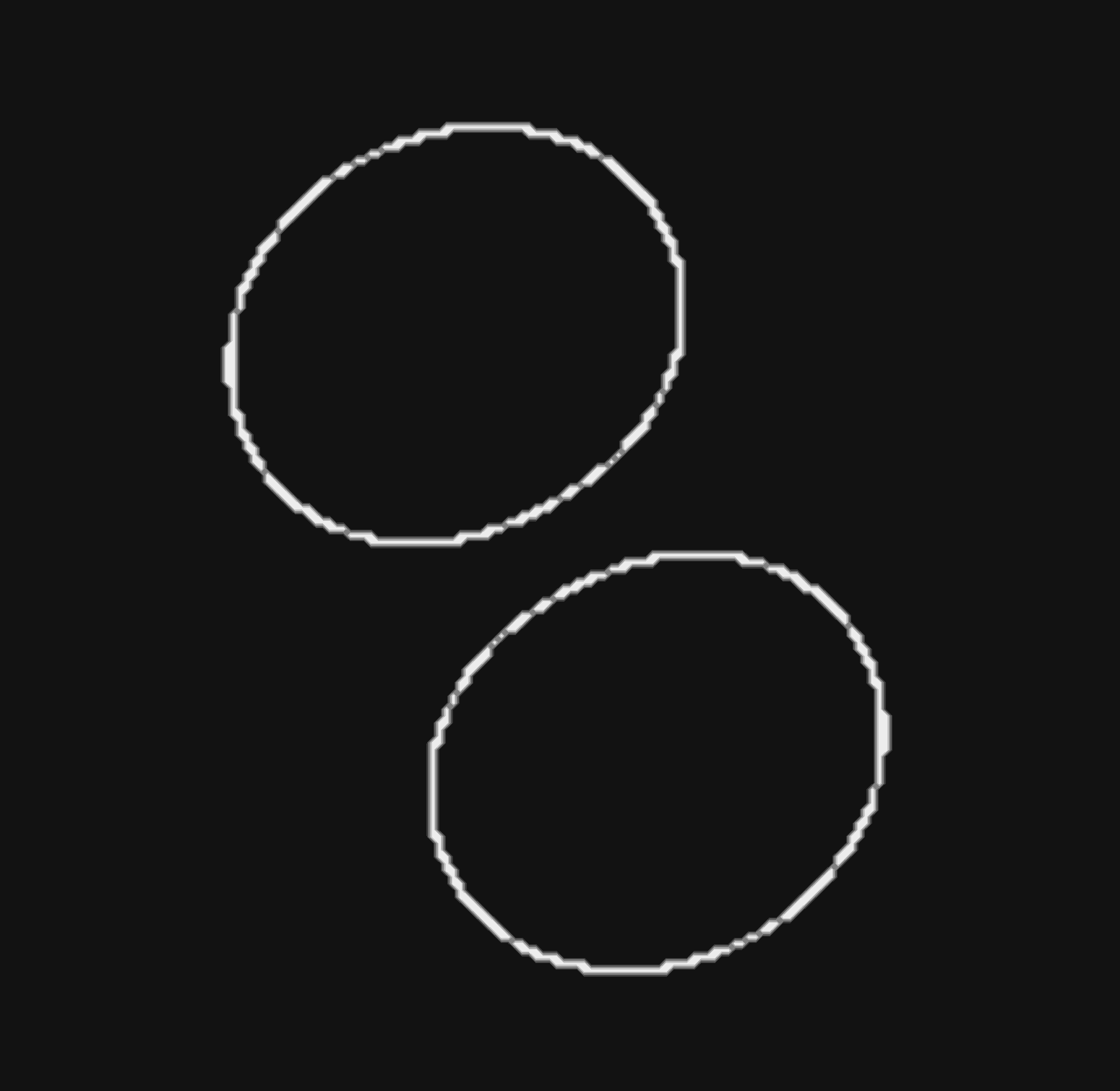}
		\includegraphics[scale = 0.18]{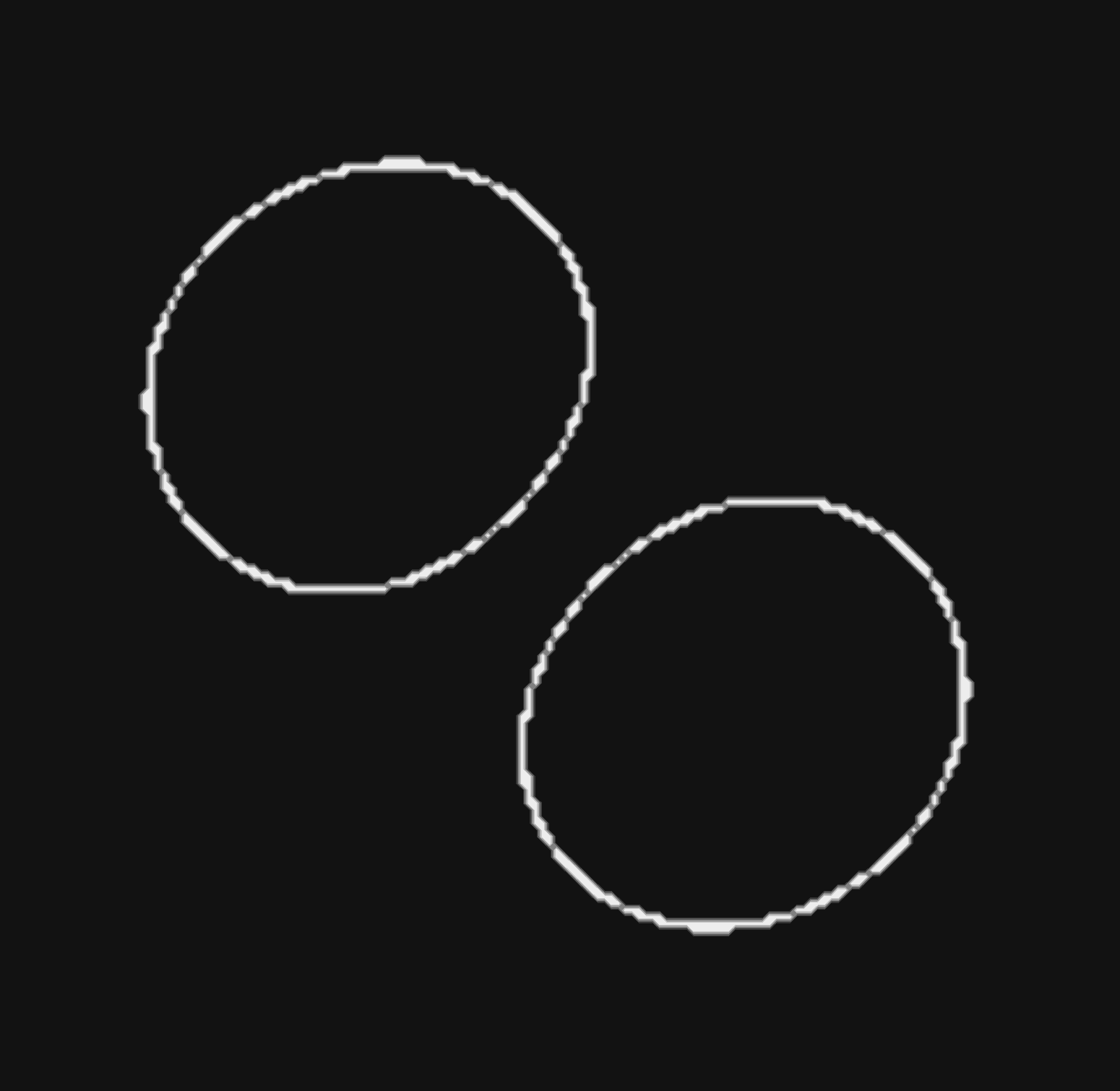}
		\includegraphics[scale = 0.18]{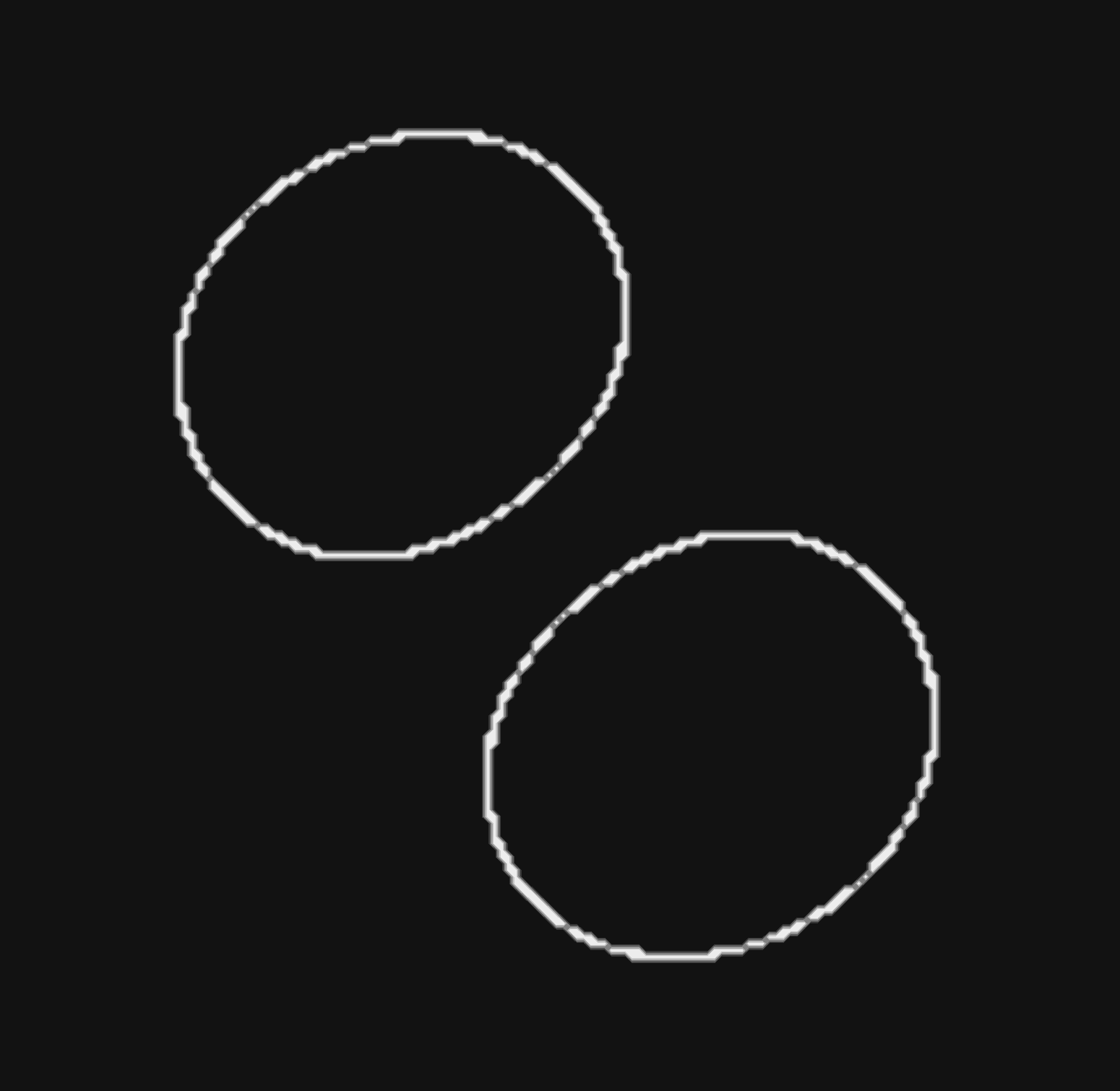}
		\includegraphics[scale = 0.18]{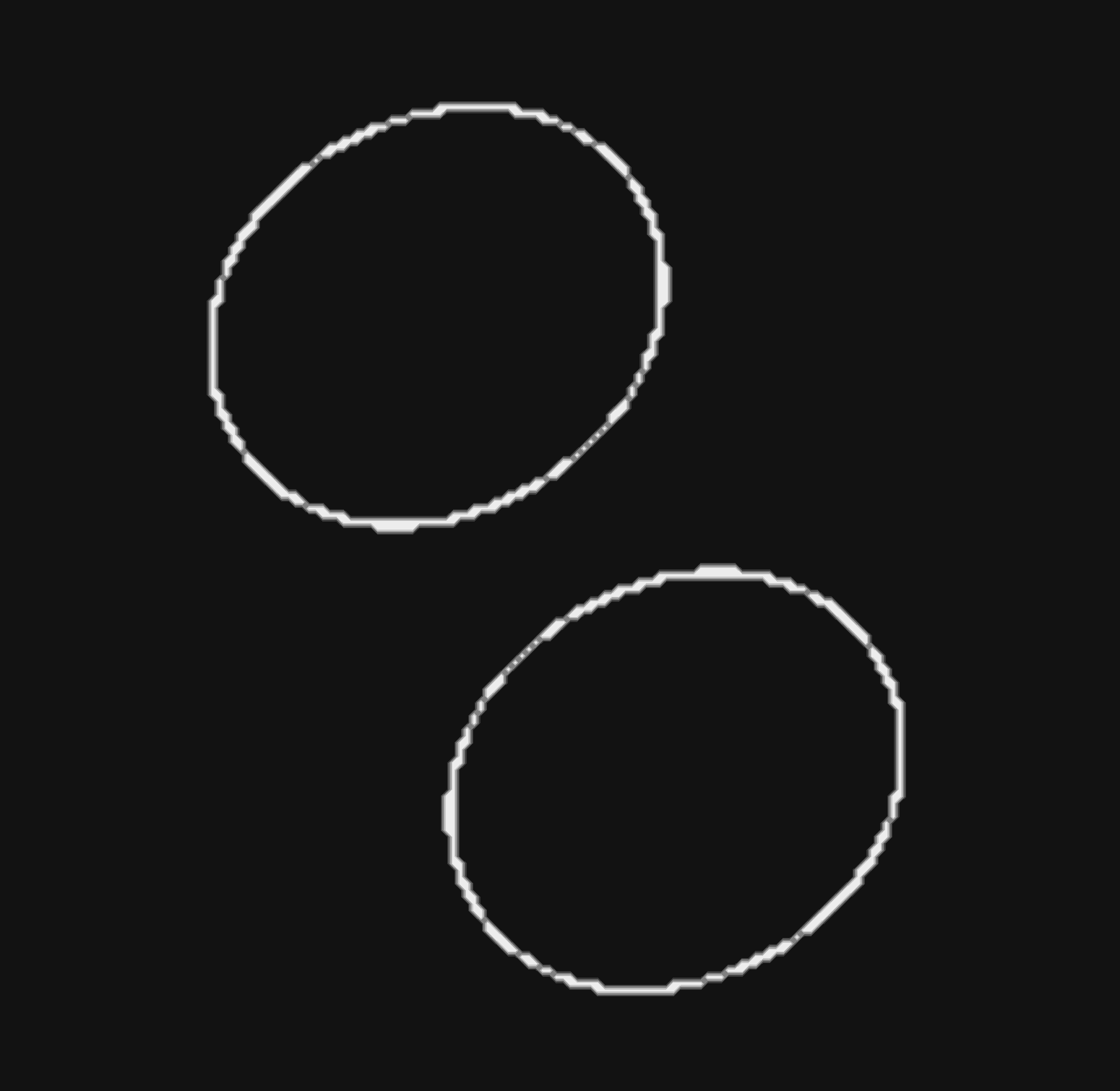}  
       \\ (c) Pass-over \\ \vspace{0.3 cm}
       
 	\caption{Sample snapshots for training from three different collision outcome cases.} 
	\label{fig:three_case_sample_snapshot}
\end{figure}

Given the focus on droplet shape and orientation for characterizing collision outcomes, we cropped the snapshots, as displayed in Figure~\ref{fig:cropping}), in a manner that encompasses both droplets during the collision. This decision aims to streamline and reduce the input dimension during training to save memory and time. Although the original simulation images were in color, we converted them into a binary grayscale format. By doing so, we effectively isolated and preserved only the droplet interface, allowing us to capture the essential shape information. This transformation simplifies the dataset by reducing color complexity, which is not essential for identifying the shape features. As a result, the model can focus more efficiently on extracting and learning the fundamental shape characteristics of the droplets, enhancing its ability to distinguish between different collision behaviors based on structural features alone.
For the testing purpose of the CNN model, we have gathered snapshots from cases with a similar initial offset and confinement distribution but featuring values different from those used in the training set. The test dataset we compiled comprises a total of 460 snapshots (120 snapshots from reverse-back, 200 snapshots from coalescence, and 140 snapshots from pass-over cases) each selected from cases of the 30 combinations of initial offset (ranging from $\frac{\Delta Y}{2R}$ = 0.075 to 0.975 with an increment of 0.1) and confinement ($\frac{2R}{H}$ = 0.305, 0.345, and 0.385). A sample set of snapshots during collision but from three different collision outcome scenarios are displayed in Figure~\ref{fig:three_case_sample_snapshot}. These snapshots offer a concise overview of the subtle differences in droplet shapes and positions during collisions, which ultimately result in three distinct outcomes. 

\begin{figure}[H]
	\centering

		\includegraphics[scale = 0.1]{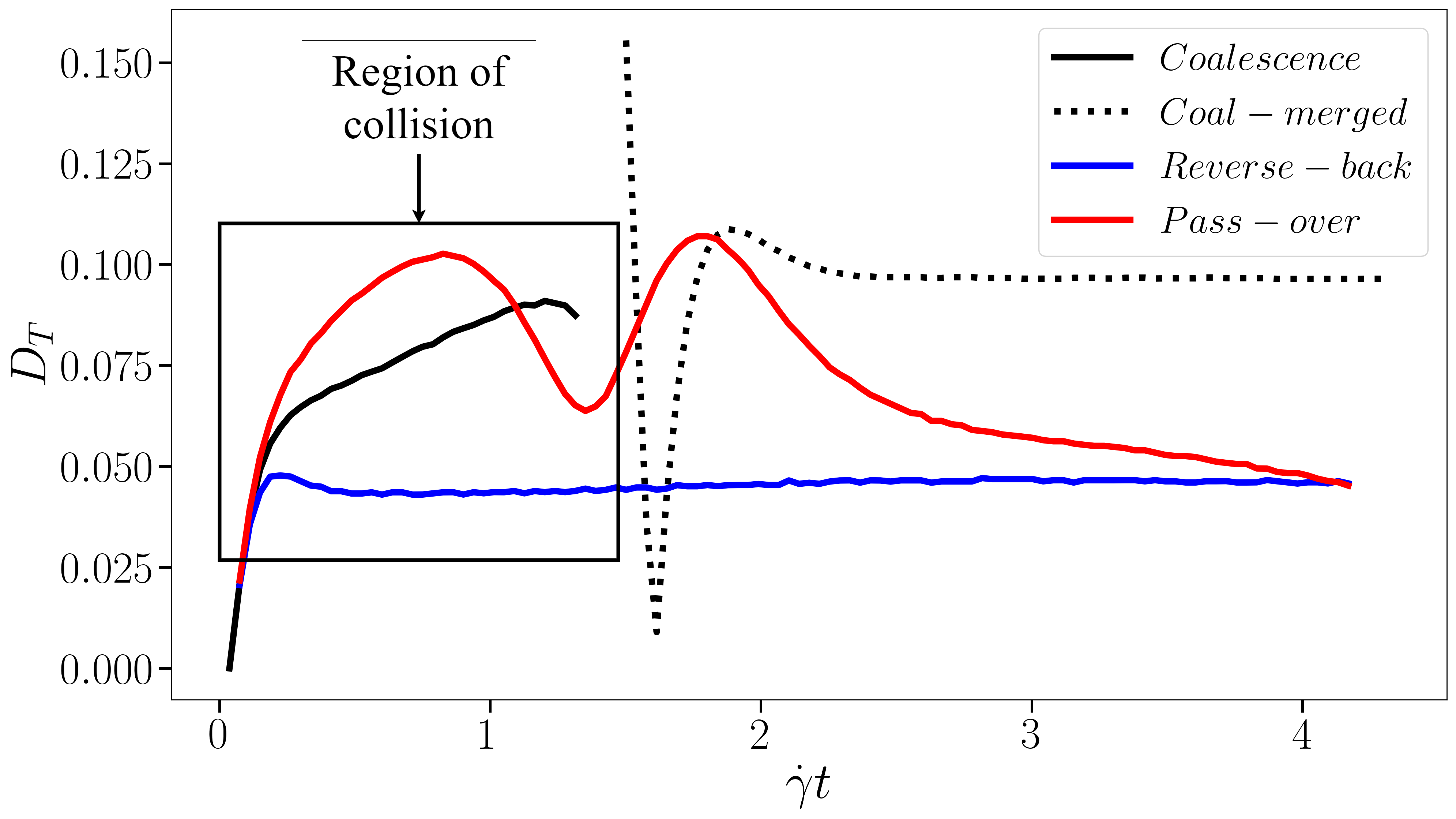} \\ \vspace{0.3cm}
		
	\caption{Taylor deformation of droplets over time for the three cases shown in Figure~\ref{fig:demo_three_different_cases}, resulting in three distinct outcomes: coalescence, reverse-back, and pass-over.}
	\label{fig:Deformation_sample}
\end{figure}

To illustrate the trends in deformation behavior arising from three distinct case scenarios considered in Figure~\ref{fig:demo_three_different_cases}, we present the Taylor deformation ($D_T = \frac{A-B}{A+B}$, where A and B represent the major and minor axes of deformed droplets, respectively) for each scenario and demonstrate in Figure~\ref{fig:Deformation_sample}. The cases include reverse-back, coalescence, and pass-over. For the coalescence scenario, we specifically highlight the deformation of the merged droplet using a dotted line, representing the post-coalescence state. Additionally, a black square in the plot indicates the collision region applicable to all three cases, serving as a visual reference point. Notably, the analysis reveals that each collision scenario exhibits unique variations in deformation characteristics. These differences underscore the importance of droplet shape, as they encapsulate crucial information that can be leveraged to predict the final outcomes of these interactions. Thus, the deformation behaviors not only reflect the dynamics of the collision but also provide insights into the underlying physical processes at play.

The dataset used in this study consists of snapshots capturing droplet shapes during collision events. Each snapshot is represented as a PNG image file, with a cropped dimension of $164 \times 164$ pixels,as discussed in Sec. B(1). We arranged the images in sequence, allocating the first $N_{train} = 3010$ snapshots for training and the final $N_{test} = 460$ snapshots for testing.
Each snapshot $\textbf{X}_i$ is represented as a two-dimensional matrix of pixel values (either 0 or 1), denoted as $\textbf{ X}_i \in \mathbb{R}^{H \times W}$ where $H$ and $W$ signify the height and width of the image, respectively.

\subsection{Model architecture}

\begin{figure}[H]
	\centering

		\includegraphics[scale = 0.6]{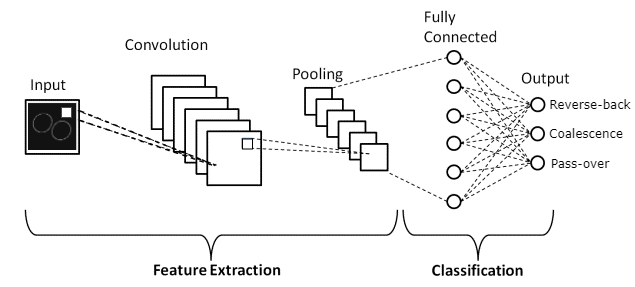} \\ \vspace{0.3cm}
		
	\caption{General architecture of the developed CNN model for classification of collision outcomes.}
	\label{fig:general_cnn_architecture}
\end{figure}

The CNN architecture is carefully crafted to capture hierarchical feature extraction from input images, leading to strong classification performance. Figure~\ref{fig:general_cnn_architecture} demonstrates the general CNN architecture framework, particularly fitted to address our classification issue. 

The entire modeling process can be subdivided into two sections as 1) Feature extraction and 2) Classification. The detailed architectural blueprint of the model we have applied is delineated as follows:

\begin{enumerate}
\item \textbf{Convolutional layer:} 
The foundational layer of the CNN architecture, a convolutional operation:
\begin{equation}
    \mathbf{Z}_i^{\left [ 1 \right ]} = \textrm{Conv}\left ( \mathbf{X}_i , \mathbf{W}^{\left [ 1 \right ]}\right) + b^{\left [ 1 \right ]},
\end{equation}
 
which conducts the spatial filtering of input images using $F$ learnable filters of size $K \times K$. Herein, $\textbf{Z}_i^{\left [ 1 \right ]}$ signifies the resultant feature maps, $\textbf{W}^{\left [ 1 \right ]}$ represents the convolutional filter weights, and $b^{\left [ 1 \right ]}$ denotes the bias vector.

\item \textbf{Activation function:} 
The rectified linear unit (ReLU) activation function is applied element-wise to introduce non-linearity: 
\begin{equation}
    \textbf{A}_i^{\left [ 1 \right ]} = \textrm{ReLU}\left ( \textbf{Z}_i^{\left [ 1 \right ]} \right )
\end{equation}

\item \textbf{Average pooling layer:} 
An average pooling operation is applied to downsample the feature maps, reducing their spatial dimensions: 
\begin{equation}
    \textbf{P}_i^{\left [ 1 \right ]} = \textrm{AvgPool}\left ( \textbf{A}_i^{\left [ 1 \right ]}, \textrm{pool\_size} \right )
\end{equation}

\item \textbf{Flatten layer:}
The output of the pooling layer is flattened into a one-dimensional vector: 
\begin{equation}
     \textbf{F}_i^{\left [ 1 \right ]}= \textrm{Flatten}\left ( \textbf{P}_i^{\left [ 1 \right ]} \right )
\end{equation}

\item \textbf{Fully connected layer: } A dense layer with $N_{classes}$ (= number of total classes) units is utilized as the output layer: 
\begin{equation}
    \textbf{Z}_i^{\left [ 2 \right ]}= \textbf{F}_i^{\left [ 1 \right ]} \times \textbf{W}^{\left [ 2 \right ]} + b^{\left [ 2 \right ]}
\end{equation}

\item \textbf{Activation function:}
Finally, the softmax activation function is applied to compute the probability distribution over the classes: 
\begin{equation}
    \hat{\textbf{Y}_i}=\textrm{Softmax}\left ( \textbf{Z}_i^{\left [ 2 \right ]} \right ).
\end{equation}

\end{enumerate}

\subsection{Model training}

The compiled CNN model is trained using the training dataset, consisting of labeled images and their corresponding collision outcomes. During training, the model is optimized to minimize the categorical cross-entropy loss function: 
 \begin{equation}
     \textbf{J}\left ( \textbf{W}^{\left [ 1 \right ]}, b^{\left [ 1 \right ]}, \textbf{W}^{\left [ 2 \right ]}, b^{\left [ 2 \right ]} \right ) = - \frac{1}{N_{train}} \sum_{i=1}^{N_{train}} \sum_{c=1}^{N_{classes}} \textbf{Y}_{i,c} \textrm{log}\left ( \hat{\textbf{Y}}_{i,c} \right ),
 \end{equation}
where, $\textbf{Y}_{i,c}$ represents the ground truth label for image $i$ and class $c$, and $ \hat{\textbf{Y}}_{i,c}$ denotes the predicted probability for class $c$ given image $i$. The model parameters (weights and biases) are updated iteratively using backpropagation and one of the optimization methods to minimize the loss function.

\subsection{Evaluation}

After training, the model is evaluated on the testing dataset, which consists of unseen images not encountered during training. The model's performance is assessed based on metrics such as accuracy and loss, computed as follows: 
\begin{equation}
    \textrm{\textbf{Accuracy}} = \frac{\textrm{Number of correctly classified samples}}{\textrm{Total number of samples}},
\end{equation}

\begin{equation}
    \textbf{\textrm{Loss}} = \textbf{J}\left ( \textbf{W}^{\left [ 1 \right ]}, b^{\left [ 1 \right ]}, \textbf{W}^{\left [ 2 \right ]}, b^{\left [ 2 \right ]} \right ) .
\end{equation}

\subsection{Training procedure}

The entire training procedure, including data loading, model compilation, training, and evaluation, is implemented using Python programming language and the Keras deep learning framework. The training process is executed on hardware equipped with suitable computational resources to facilitate efficient model optimization and convergence. By following this methodology, the study aims to develop a CNN-based predictive model capable of accurately classifying droplet collision outcomes based on shape analysis, thereby advancing our understanding and control of complex fluid dynamics phenomena.

\section{\label{sec:results}Results and Discussion}

\subsection{Learning rate tuning}

In CNN training, the learning rate is a crucial hyperparameter that controls the step size of each iteration as the model moves toward minimizing the loss function. It determines how quickly or slowly a model learns. A learning rate that is too high can cause the model to converge too quickly to a suboptimal solution, while a learning rate that is too low can result in a long training process that might get stuck in local minima.
The learning rate directly influences the efficiency and effectiveness of the training process. An optimal learning rate ensures that the model converges at a reasonable pace while finding the best possible weights for the network. Adjusting the learning rate is often necessary to achieve the best performance from the model. Small adjustments can have significant impacts on the final accuracy and generalization capability of the model.

\begin{figure}[H]
	\centering

	\includegraphics[scale = 0.40]{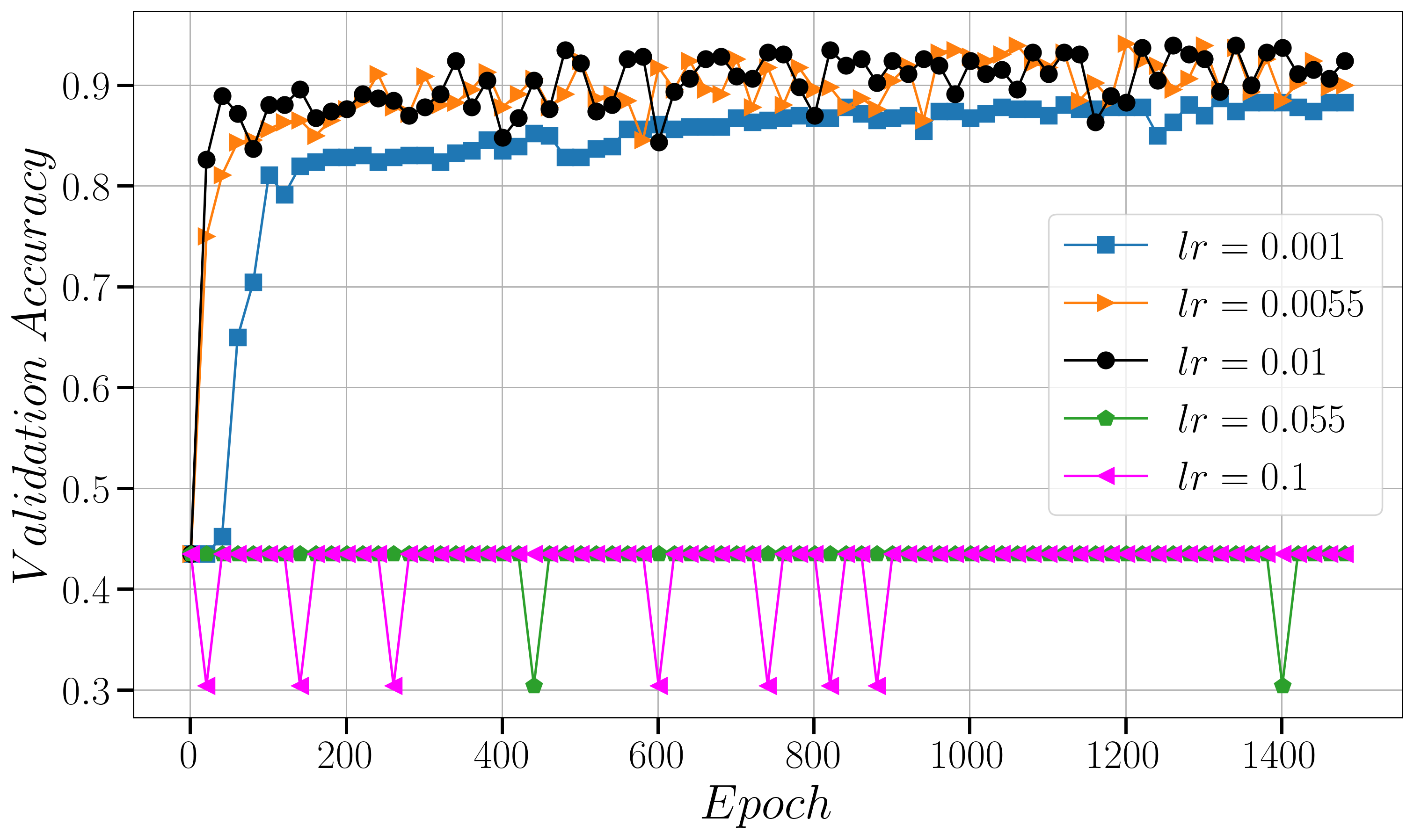} 
 \\ (a) Overall accuracy \\ \vspace{0.3 cm}
 \includegraphics[scale = 0.40]{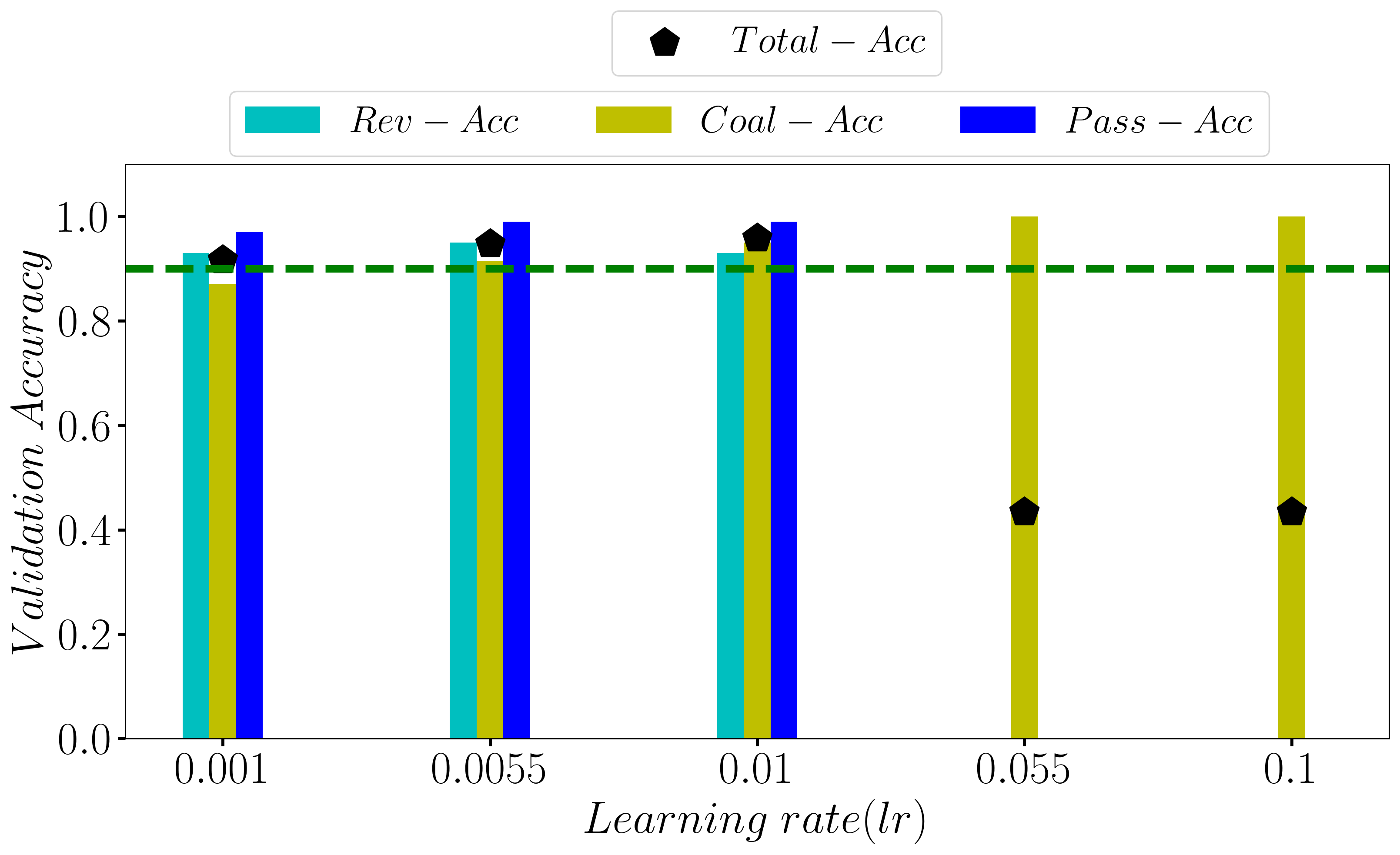} 
 \\ (b) Class-wise accuracy
		
	\caption{Impact of learning rate on CNN accuracy for droplet collision prediction. (a) Overall accuracy evaluations with learning rates of $lr=0.001$, $0.0055$, $0.01$, $0.055$, and $0.1$ over 1500 epochs. (b) Class-wise prediction accuracy (ACC) for different learning rates: the bar plot illustrates the prediction accuracies of the CNN model for three droplet collision outcome classes across various learning rates, with the total accuracy for each learning rate indicated by a black star. The green dashed line indicates the accuracy level of $0.90$. The model architecture included an initial convolutional layer with 10 filters of size $14 \times 14$, ReLU activation, followed by average pooling, and a final softmax layer.}
	\label{fig:lr_tuning}
\end{figure}

To determine the optimal learning rate for the CNN model predicting droplet collision outcomes, we experimented with different learning rates. The results, assessed over 1500 epochs as demonstrated in Figure~\ref{fig:lr_tuning}(a). While our primary objective was to tune the learning rate, we selected other hyperparameters and methods as initial trial settings to establish a baseline. We employed the Root Mean Square Propagation (RMSprop) optimizer, which is effective in managing the non-stationary objectives often encountered in deep learning~\cite{goodfellow2016deep}. The model architecture included an initial convolutional layer with 10 filters of size $14 \times 14$ to extract spatial features from the images. To ensure regularization and mitigate overfitting, we applied a weight decay of $10^{-6}$. The first layer used the ReLU activation function for its ability to address the vanishing gradient problem. This was followed by an average pooling layer to downsample the spatial dimensions of the feature maps, reducing computational complexity. The final layer utilized a softmax activation function to generate a probability distribution over the three classes of collision outcomes, facilitating multi-class classification. These initial choices enabled us to focus on optimizing the learning rate, providing a practical starting point for further refinement.

The selection of an appropriate learning rate is critical for the effective training of a CNN model. In this study, we explored the impact of five different learning rates ($l_r=0.001$, $0.0055$, $0.01$, $0.055$, and $0.1$) on the model's ability to predict droplet collision outcomes, namely reverse-back, coalescence, and pass-over. The overall accuracy and class-wise prediction accuracies were evaluated for each learning rate setting. The results of our computational experiments showed that a learning rate of 0.01 achieved the highest total accuracy at 0.958. This was followed by a learning rate of 0.0055, which achieved a total accuracy of 0.947, and a learning rate of 0.001 with 0.917. The learning rates of 0.1 and 0.055 both resulted in significantly lower accuracies of 0.434. This indicates that the learning rate of $0.01$ provides an optimal balance, allowing the model to effectively navigate the loss landscape and reach a solution that generalizes well to new data. 

		
To gain deeper insights, we analyzed the prediction accuracies for each of the three collision outcome classes individually, asb illustrated in Figure~\ref{fig:lr_tuning}(b). From the class-wise analysis, it is evident that the learning rate of 0.01 not only achieves the highest overall accuracy but also maintains superior performance across all three classes. For the reverse-back cases, the learning rates of $0.001$ and $0.01$ performed well with accuracies of $0.93$ each, while the learning rate of $0.0055$ performed even better at $0.95$. Learning rates of $0.1$ and $0.055$ failed completely, indicating instability. On the other hand, to classify coalescence cases the learning rate of $0.01$ showed significant improvement ($0.95$) over $0.001$ ($0.87$), while $l_r=0.0055$ also performed well at $0.915$. The learning rates of $0.1$ and $0.055$ achieved perfect accuracy ($1.00$) but at the expense of performance in other classes. Finally, if we take a look at the prediction accuracy of the pass-over cases, then it can be noticed that the learning rates of $0.01$ and $0.0055$ achieved the highest accuracy ($0.99$), slightly outperforming $l_r=0.001$ ($0.97$). Learning rates of $0.1$ and $0.055$ again performed poorly ($0.00$). These findings are visually represented in Figure~\ref{fig:lr_tuning}(b), where class-wise accuracies are shown as color bars for each learning rate, and total accuracies are marked with black stars. The figure clearly demonstrates that the learning rate of $0.01$ provides the most balanced and robust performance across all classes. Consequently, we selected a learning rate of 0.01 for further analyses and training in this study.

\subsection{Optimizer performance analysis}

\begin{figure}[H]
	\centering

	\includegraphics[scale = 0.40]{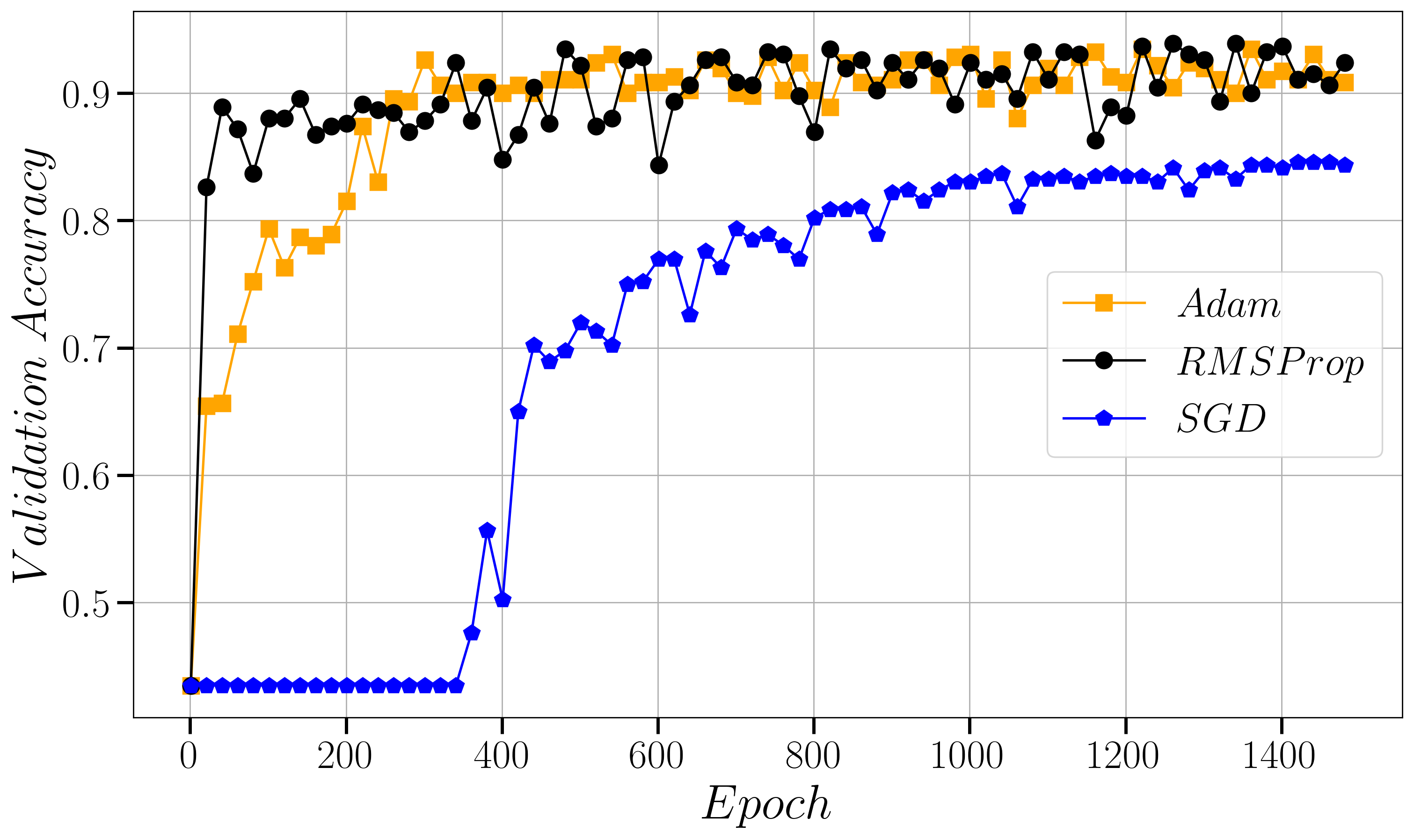}
 \\ (a) Overall accuracy \\ \vspace{0.3 cm}
\includegraphics[scale = 0.40]{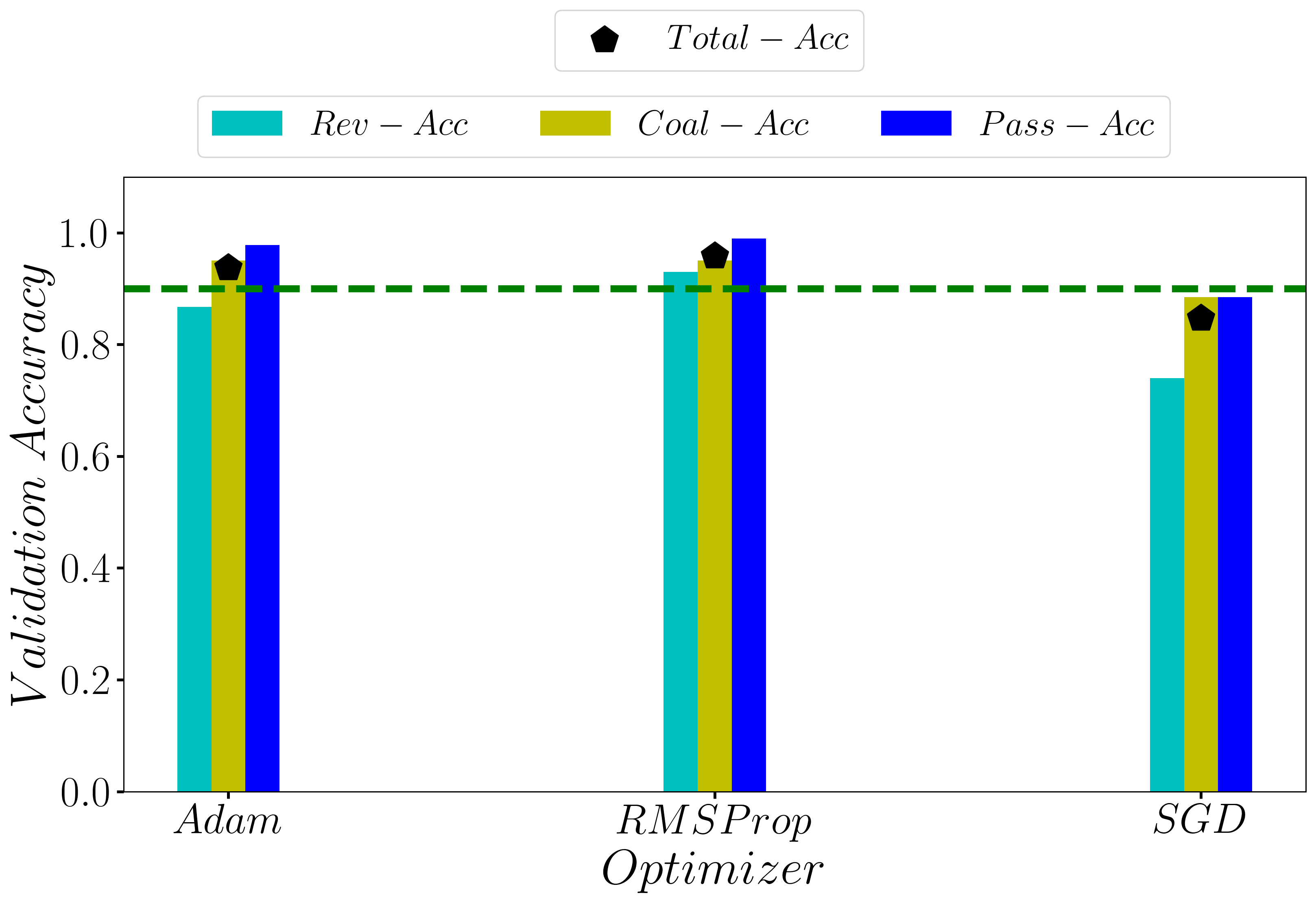}
\\ (b) Class-wise accuracy \\
		
	\caption{Performance of different optimizers on CNN accuracy for droplet collision prediction with learning rates of $0.01$ over 1500 epochs. (a) Overall accuracy for different optimizers. (b) Class-wise prediction performance: bar plot illustrates the prediction accuracies (ACC) of the CNN model for three classes of droplet collision outcomes (i.e., reverse-back (Rev), coalescence (Coal), pass-over (Pass)), with the total accuracy achieved for each optimizer indicated by black stars. The green dashed line indicates the accuracy level of $0.90$. The model architecture included an initial convolutional layer with 10 filters of size $14 \times 14$, ReLU activation, followed by average pooling, and a final softmax layer.}
	\label{fig:optimizer_epochs}
\end{figure}

The performance of a CNN model is heavily influenced by the choice of optimizer, which is crucial for model training and convergence. An optimizer is an algorithm that adjusts neural network weights to minimize the loss function, guiding the model toward an optimal solution. Different optimizers have unique strategies for updating weights, affecting the learning process significantly. In predicting droplet collision outcomes from shape information, selecting the right optimizer is vital as it directly affects convergence rate, stability, and accuracy. Choosing the right optimizer is critical for several reasons. The optimizer used directly affects how quickly a model converges, and faster convergence improves training efficiency, especially with large datasets or complex models. Additionally, the stability of the optimization process is paramount. Optimizers like RMSProp and  Adaptive Moment Estimation (Adam) provide more stable updates, reducing the risk of the model getting stuck in local minima or experiencing large  fluctuations in loss. Moreover, an effective optimizer helps balance minimizing training loss and achieving good generalization on unseen data, ensuring that the model performs well in real-world scenarios. Finally, optimizers that dynamically adjust the learning rate, such as RMSProp and Adam, can better handle varying scales of gradient, leading to improved performance in complex tasks like predicting droplet collision outcomes.
After identifying an optimal learning rate of $l_r=0.01$ in Sec. III(A), we evaluated the performance of three optimizers—Stochastic Gradient Descent (SGD), RMSProp, and Adam—for predicting droplet collision outcomes, while keeping other hyperparameters constant. Our aim is to understand how each
optimizer influences the model’s training dynamics, convergence behavior, and overall accuracy. The validation accuracy over the $1500$ epochs for each of the optimizer have been demonstrated in Figure~\ref{fig:optimizer_epochs}(a). 
The validation accuracy curve for RMSProp shows a steady and rapid increase, stabilizing at a high accuracy level. This indicates that RMSProp efficiently optimizes the model, achieving high accuracy with fewer epochs. On the other side, the validation accuracy curve for Adam also shows a steady increase, though it converges slightly slower than RMSProp. Adam stabilizes at a high accuracy level but slightly lower than RMSProp. In contrast, the validation accuracy curve for SGD shows a slower increase and greater fluctuations compared to RMSProp and Adam, stabilizing at a lower accuracy level, indicating less effective optimization.

The overall accuracy achieved with each optimizer, representing the best accuracy obtained over the 1500 epochs, provides an initial indication of their performance. RMSProp achieved the highest overall accuracy at $0.958$, followed by Adam at $0.937$, and SGD at $0.847$. This suggests that RMSProp is the most effective optimizer for our model, providing the best generalization to new data. 

To gain deeper insights, we analyzed the prediction accuracies for each collision outcome class individually. The results are visually represented in Figure~\ref{fig:optimizer_epochs}(b), where class-wise accuracies are shown as color bars for each optimizer, with total accuracy marked by black stars. From the class-wise analysis, it is evident that RMSProp not only achieves the highest overall accuracy but also maintains superior performance across all three classes. In classifying reverse-back cases, RMSProp achieved the highest accuracy ($0.93$), followed by Adam ($0.867$) and SGD ($0.74$). RMSProp’s effectiveness here suggests it can better capture the subtleties of reverse-back cases during collisions. In coalsecence prediction, RMSProp and Adam both performed equally well with an accuracy of $0.95$, significantly outperforming SGD ($0.885$). This indicates RMSProp and Adam’s robustness in predicting coalescence outcomes. Finally RMSProp achieved the highest accuracy ($0.99$), closely followed by Adam ($0.978$) and SGD ($0.885$), indicating RMSProp’s superior ability to predict pass-over collisions.
RMSProp’s superior performance can be attributed to its adaptive learning rate mechanism. RMSProp adjusts the learning rate dynamically for each parameter, allowing for efficient and stable training. This adaptability helps the optimizer navigate the loss landscape effectively, preventing overshooting and ensuring convergence. The ability to fine-tune the learning rate for each parameter allows RMSProp to handle the complexities and variations in the data more effectively than SGD, which uses a fixed learning rate. 

Based on the overall and class-wise accuracies, as well as the training curves, we conclude that RMSProp is the most effective optimizer for our CNN model in predicting droplet collision outcomes. Its adaptive learning rate facilitates rapid yet stable convergence, resulting in superior generalization across all classes. Adam also performs well but does not match the performance of RMSProp in this application. SGD, with its fixed learning rate, falls significantly behind in terms of both overall and class-wise accuracies.

\subsection{Optimizing filter count}

In the realm of CNN modeling, the architecture of the network plays a pivotal role in determining the model's ability to effectively learn and generalize from data. One critical aspect of this architecture is the number of filters used in the convolutional layers. Filters, also known as kernels, are essential components that detect specific features in the input data by convolving with the input image to produce feature maps. These feature maps highlight various patterns such as edges, textures, and shapes that are crucial for tasks like image recognition and classification.

When dealing with a specialized application like predicting droplet collision outcomes from shape information, the selection of the number of filters becomes even more significant. The droplet collision outcomes (reverse-back, coalescence, and pass-over) are highly dependent on subtle shape variations that occur during the collision process. Therefore, the ability of the CNN to capture these intricate details hinges on the optimal configuration of filters. Filters in CNN serve to identify and extract different levels of spatial hierarchies from the input images. At lower layers, filters typically capture basic features such as edges and corners, while higher layers capture more complex patterns and shapes. The number of filters at each layer determines the breadth of features the network can learn. A low number of filters may result in the model lacking the capacity to capture the full diversity of features present in the input data. This limitation can lead to inadequate learning of intricate shape details necessary for distinguishing between different collision outcomes. Conversely, a higher number of filters increases the network’s ability to detect a wide range of features. This increased capacity allows the model to capture finer details and more complex patterns, potentially leading to better performance. However, an excessively high number of filters can also introduce redundancy, increase computational cost, and lead to overfitting, where the model performs well on training data but poorly on unseen data.
The challenge lies in finding a balance where the number of filters is sufficient to capture the necessary features without introducing redundancy or overfitting. This balance is crucial for achieving high accuracy in predicting droplet collision outcomes.

\begin{figure}[H]
	\centering

	\includegraphics[scale = 0.40]{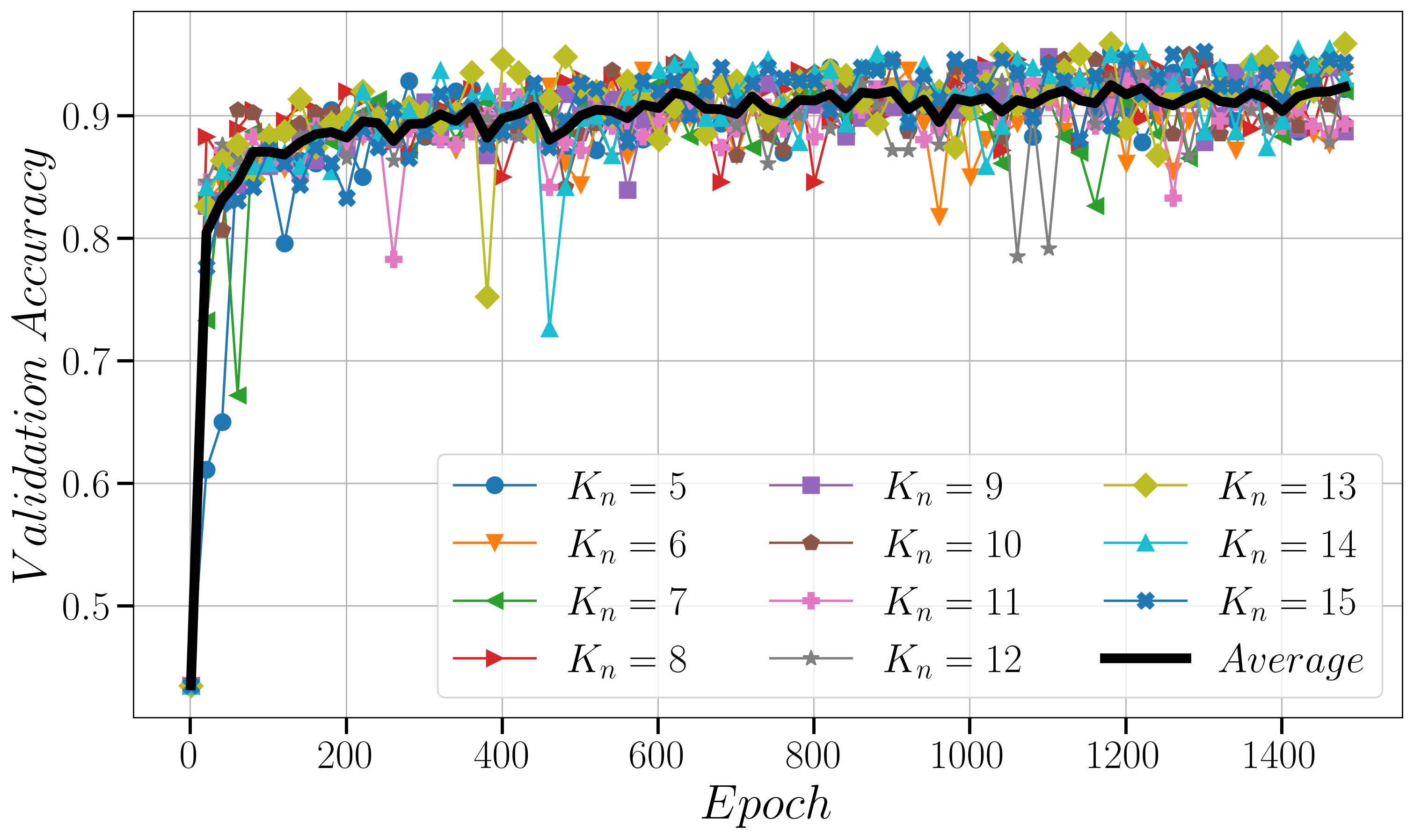}
 \\ (a) Overall accuracy \\ \vspace{0.3 cm}
 \includegraphics[scale = 0.35]{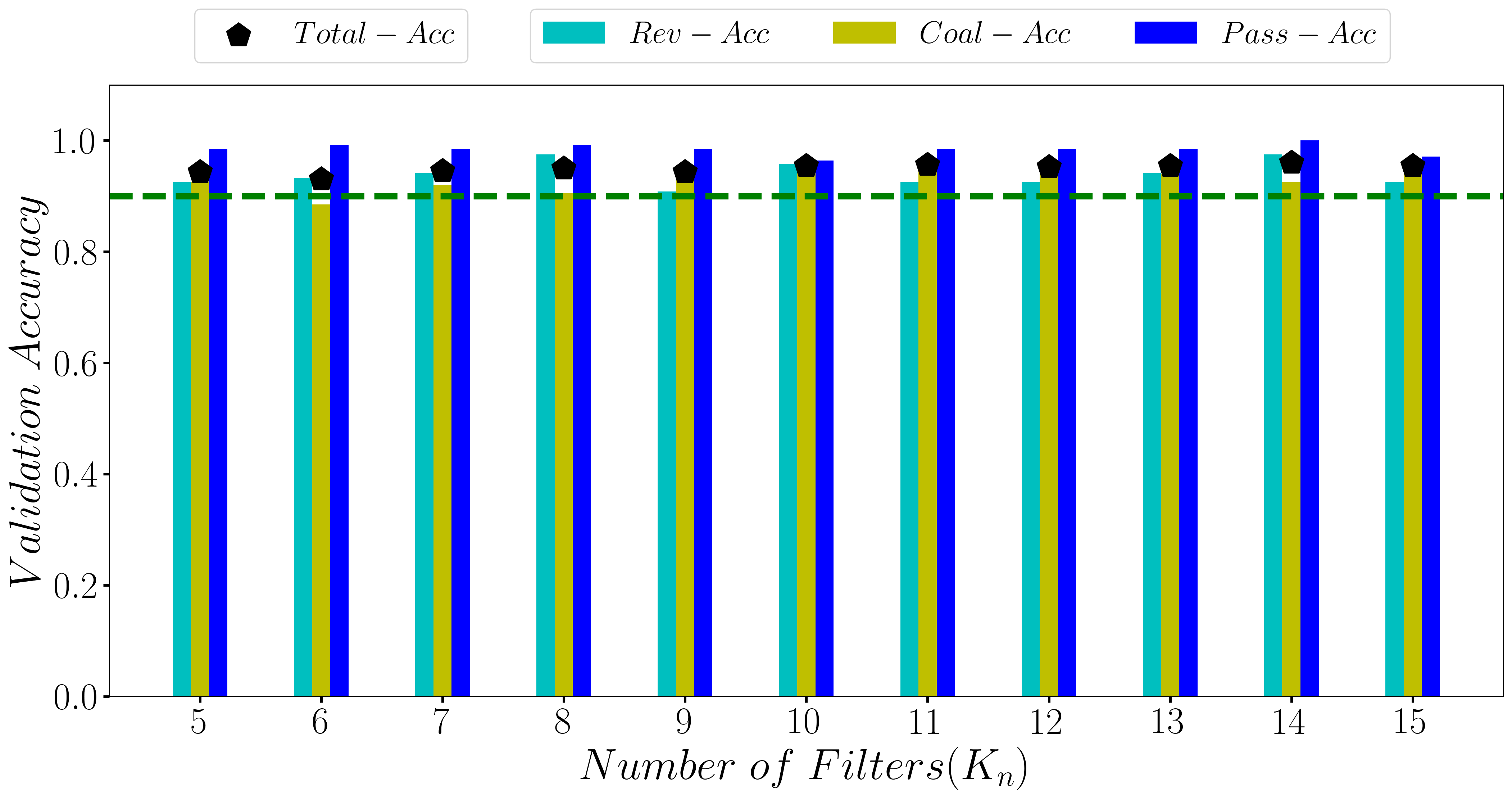}
 \\ (b) Class-wise accuracy
		
	\caption{Interplay between the number of filters ($5$ to $15$) and the CNN’s performance in predicting droplet collision outcome with learning rates of $l_r=0.01$ and RMSProp optimizer over 1500 epochs. (a) Overall accuracy for different filters. (b) Class-wise prediction performance: bar plot illustrates the prediction accuracies of the CNN model for three classes of droplet collision outcomes (reverse-back, coalescence, pass-over), with the total accuracy achieved for each optimizer indicated by black stars. The green dashed line indicates the accuracy level of $0.90$. The model architecture included an initial convolutional layer with filters of size $14 \times 14$, ReLU activation, followed by average pooling, and a final softmax layer.}
	\label{fig:filter_number_epochs}
\end{figure}

Here, we have explored the impact of changing the number of filters on the CNN’s ability to predict droplet collision outcomes. By conducting this analysis with filter numbers ranging from $5$ to $15$ and using our previously established optimal learning rate along with the RMSProp optimizer, we seek to understand how the number of filters influences the model's ability to capture shape information and accurately predict collision outcomes. The resulting validation accuracy curves, along with an averaged accuracy curve, over 1500 epochs are illustrated in Figure~\ref{fig:filter_number_epochs}(a)  for each filter number configuration, ranging from
5 to 15.

The black curve in Figure~\ref{fig:filter_number_epochs}(a)  represents the average validation accuracy across all filter number cases. The results of Figure~\ref{fig:filter_number_epochs}(a) provides valuable insights into the model’s performance and stability. The validation accuracy curves for each filter number exhibits similar overall trends, suggesting that the model’s performance is relatively consistent across different filter configurations. However, some curves demonstrate instability at certain points, reflecting fluctuations in accuracy during training. In contrast, the black curve, representing the averaged validation accuracy across all filter numbers, shows a smooth and stable trend. This indicates that, on average, the model maintains consistent performance irrespective of the specific number of filters used.

The consistency in performance across different filter numbers suggests that our CNN model is robust to variations in the number of filters. This robustness is likely due to the adaptive nature of the RMSProp optimizer and the chosen learning rate, which collectively ensure stable and effective training. The similar trends across different filter configurations also indicate that the model is capable of capturing the necessary features for predicting droplet collision outcomes regardless of the exact number of filters. Despite the overall consistency, some configurations exhibit instability at specific epochs. This instability could be attributed to several factors. Variations in the learning dynamics for different filter numbers can lead to temporary fluctuations in accuracy. This is especially true for smaller filter numbers, where the model may struggle to capture all relevant features, resulting in unstable performance. Additionally, for configurations with very few filters, the model might underfit, failing to capture complex patterns in the data. Conversely, configurations with many filters could lead to overfitting, where the model becomes too tailored to the training data, resulting in fluctuations in validation accuracy.

In addition to analyzing the overall validation accuracy, we conducted a detailed assessment of class-wise prediction accuracy for each filter number configuration, ranging from 5 to 15 filters. The results are summarized in Figure~\ref{fig:filter_number_epochs}(b), which shows class-wise prediction accuracy as colored bars for different filter numbers, along with the total prediction accuracy indicated by black stars. 
The reverse-back class in Figure~\ref{fig:filter_number_epochs}(b) shows variable performance across different filter numbers, with some configurations exhibiting lower accuracy, indicating difficulty in capturing the specific features associated with the reverse-back outcome. In contrast, the accuracy for predicting the coalescence outcome is generally higher and more consistent across different filter numbers, suggesting that the model can effectively capture the features pertinent to this class. Similarly, the pass-over class prediction accuracy remains relatively stable across different configurations, although certain filter numbers, such as 14, demonstrate superior performance, indicating a better ability to identify this outcome accurately.



		

Among the various filter configurations tested, the configuration with 14 filters stands out for several reasons. First, it provides balanced performance, delivering high prediction accuracy across all three collision outcome classes. This balance is crucial for ensuring that the model does not favor one class over others, leading to a more robust and reliable predictor. Furthermore, the overall prediction accuracy for the filter number 14 configuration is the highest among all configurations, achieving a total accuracy of 0.96. This indicates that this configuration captures the essential features needed for accurate prediction more effectively than other configurations. Lastly, despite some instabilities observed in other filter configurations, the filter number 14 configuration shows consistent performance across different epochs, suggesting stable learning dynamics.

In conclusion, our analysis of varying the number of filters in the CNN model reveals that the model maintains consistent and robust performance across different configurations. The observed instabilities at specific epochs in certain configurations highlight areas for potential improvement. However, the overall stability and similarity in performance trends underscore the model’s resilience and effectiveness in predicting droplet collision outcomes. Importantly, among the different filter configurations tested, the configuration with 14 filters demonstrated the highest total prediction accuracy of $0.96$. Also the results from the class-wise prediction accuracy analysis highlight the importance of selecting an optimal number of filters. This configuration provided the best balance between capturing the necessary features for accurate predictions and maintaining stability during training. Therefore, we have decided to use $14$ filters for the rest of the analysis. This decision is driven by the need to maximize the model’s predictive accuracy while ensuring computational efficiency and stability.

\subsection{Optimizing filter dimension}

In our ongoing effort to optimize the CNN model for predicting droplet collision outcomes, the next critical parameter to examine is the filter size. The filter size in a CNN defines the dimensions of the kernel used in convolution operations, which play a fundamental role in feature extraction. By exploring filter sizes ranging from $4 \times 4$ to $20 \times 20$, we aim to understand how different scales of feature extraction influence the model's ability to capture shape information and predict collision outcomes accurately.

Filter size is a crucial hyperparameter in CNNs because it determines the receptive field of each convolutional layer. The receptive field refers to the region of the input image that a particular filter can "see" or process. Different filter sizes capture features at varying scales, where smaller filters offer several benefits. They are adept at capturing fine-grained details and local patterns in the input image, making them beneficial for detecting small edges, textures, and minute shape variations that are critical in distinguishing subtle differences between droplet collision outcomes. Additionally, smaller filters result in fewer parameters, which can reduce the computational load and the risk of overfitting, especially when combined with appropriate regularization techniques.
In a comparable way, larger filters bring their own advantages. They cover a more extensive area of the input image, making them suitable for capturing broader and more abstract features. These filters are effective in recognizing larger patterns and contextual information that may be essential for understanding the overall shape and interaction of colliding droplets. However, larger filters also have more parameters, which can lead to increased computational demands and potential overfitting if not managed correctly. Nevertheless, they can enhance the model’s capacity to learn complex representations.
Choosing the optimal filter size involves balancing the ability to capture relevant features with the complexity and efficiency of the model. The primary objective of analyzing different filter sizes is to identify the configuration that maximizes the model's predictive accuracy and robustness. This analysis will evaluate performance by assessing the model’s performance across different filter sizes to determine the optimal configuration for feature extraction. It will also provide insights into how varying filter sizes affect the model’s ability to capture and learn relevant features from droplet collision images. Furthermore, the analysis aims to optimize accuracy, striving to achieve the highest possible accuracy for each collision outcome class and overall prediction accuracy, thereby ensuring the model’s reliability and applicability in real-world scenarios.

\begin{figure}[H]
	\centering

	\includegraphics[scale = 0.30]{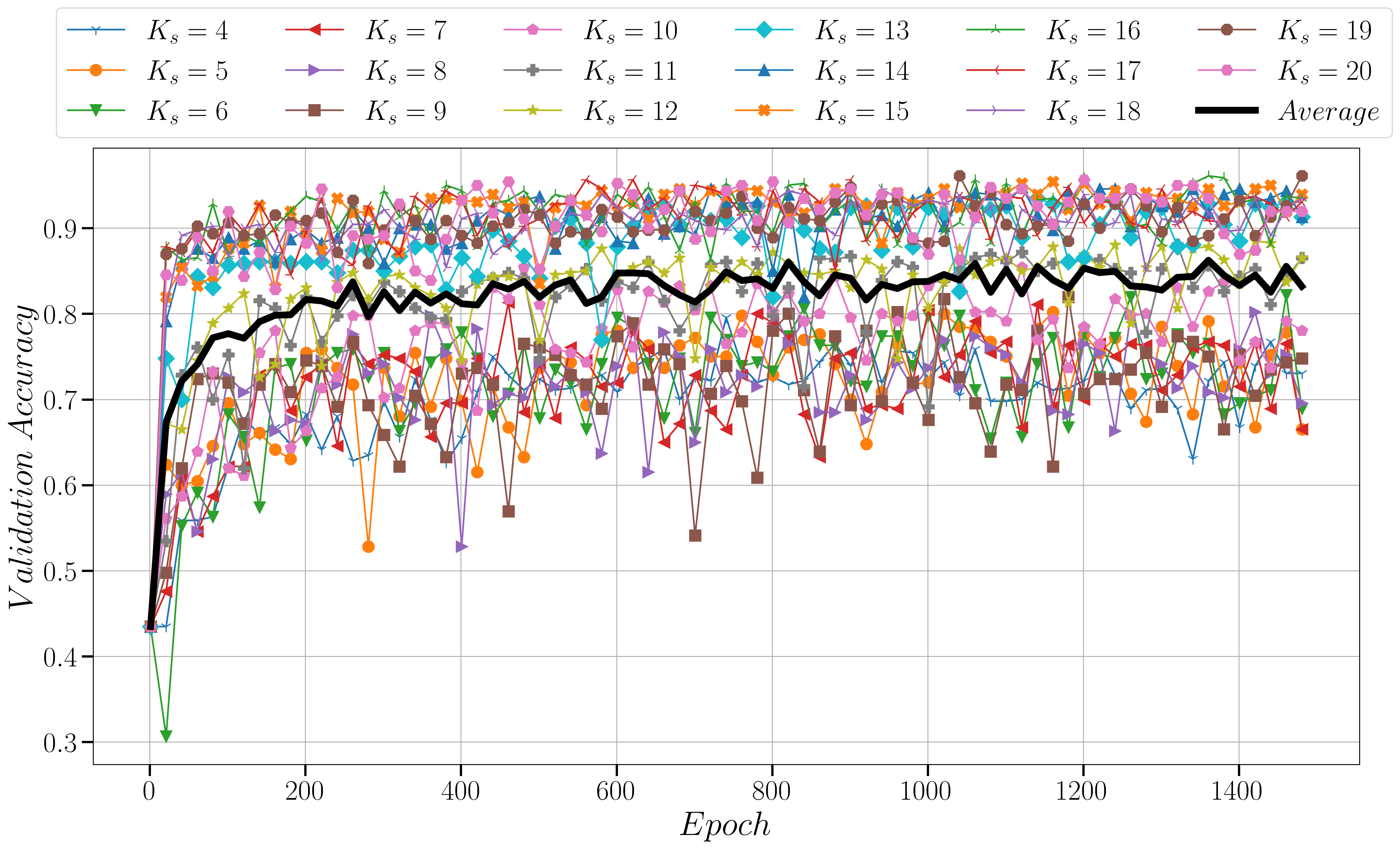}
 \\ (a) Overall accuracy \\ \vspace{0.3 cm}
 \includegraphics[scale = 0.20]{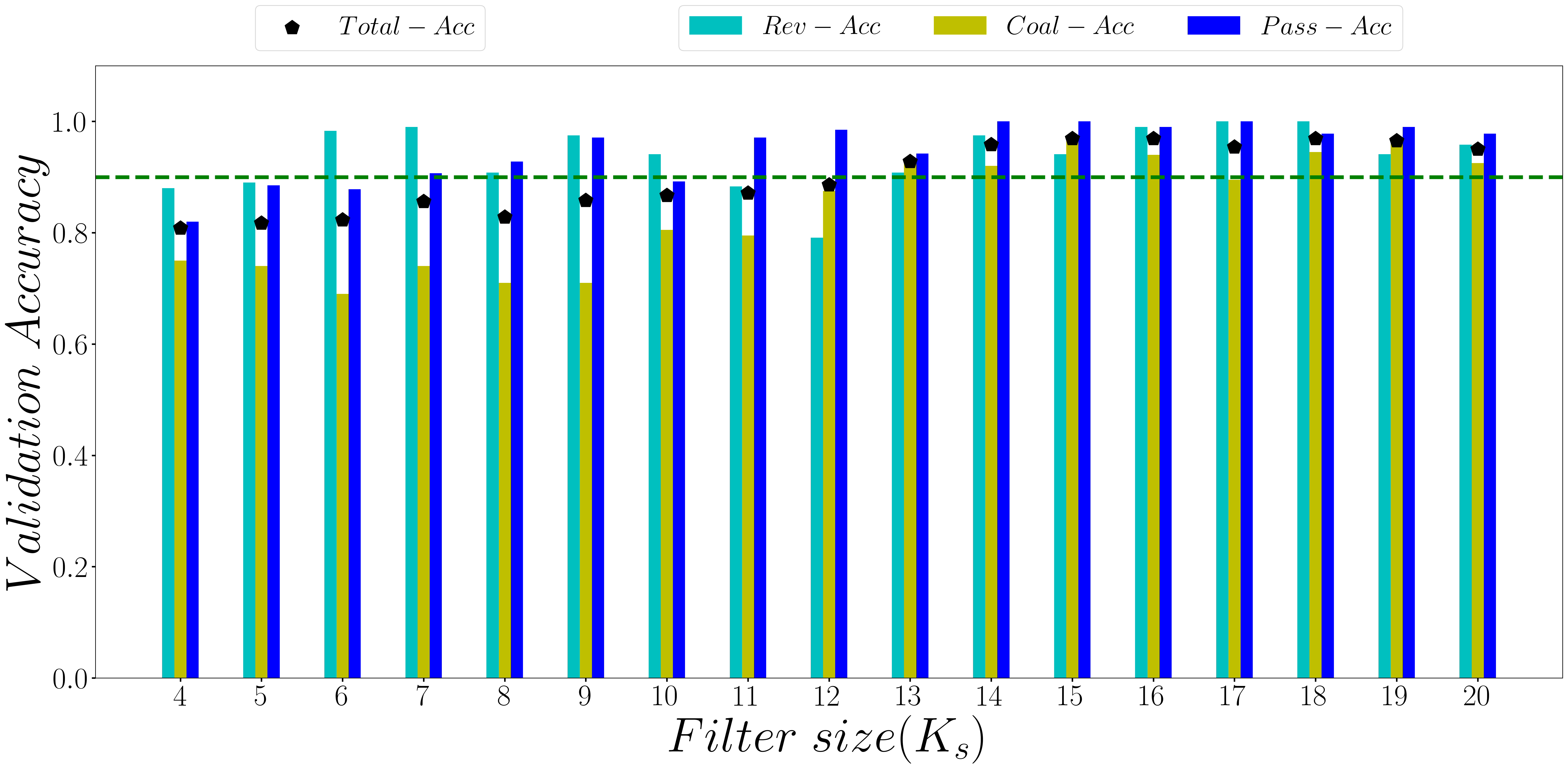}
 \\
		
	\caption{Validation accuracy over 1500 epochs for varying filter sizes ($4 \times 4$ to $20 \times 20$).  Evaluations with learning rates of $l_r=0.01$ and RMSProp optimizer. The model architecture included an initial convolutional layer with 10 filters of size $14 \times 14$, ReLU activation, followed by average pooling, and a final softmax layer. (a) Overall prediction accuracy with the black curve presenting the averaged accuracy across all filter sizes. (b) Class-wise prediction accuracy for different filter sizes for the reverse-back, coalescence, and pass-over classes using bars. The total accuracy achieved for each optimizer is indicated by a black star. The green dashed line indicates the accuracy level of $0.90$.}
	\label{fig:filter_size_epochs}
\end{figure}

By systematically exploring filter sizes from $4 \times 4$ to $20 \times 20$, we aim to refine our CNN model further, building on the insights gained from previous analyses of learning rate, optimizer selection, and the number of filters. This comprehensive approach will help us develop a highly optimized model for droplet collision outcome prediction. The results with varying filter sizes are presented in Figure~\ref{fig:filter_size_epochs}(a), which illustrates the validation accuracy trends and the averaged accuracy across all filter sizes.

The analysis reveals several key observations. First, filters with smaller dimensions (e.g., $4 \times 4, 6 \times 6$, etc.) exhibit significant instability and fail to achieve validation accuracy beyond 0.90. This underperformance can be attributed to the spatial characteristics of the droplet images; our droplets are centrally positioned, leading smaller filters to predominantly convolve around the peripheral empty spaces in the early layers, thereby failing to extract meaningful features effectively. This limitation highlights the inadequacy of small filters in capturing the essential details required for accurate predictions.

In contrast, starting from filter sizes of $13 \times 13$ and onwards, the validation accuracy consistently exceeds 0.90, demonstrating improved stability and performance. Larger filters possess a broader receptive field, allowing them to cover more significant portions of the image and effectively capture the essential features of the droplet collisions. This enhanced feature extraction capability explains the superior performance and stability of larger filter sizes. Among the tested filter sizes, the configuration with $18 \times 18$ filters emerged as the most balanced option, achieving high overall validation accuracy while also exhibiting consistent performance across different epochs. The averaged accuracy curve further supports this finding, indicating that larger filters generally provide better and more stable performance.


		
Figure~\ref{fig:filter_size_epochs}(b) shows the overall validation accuracy analysis as a function of filter sizes for each collision outcome class (i.e., reverse-back, coalescence, pass-over), revealing a trend similar to that observed in the overall accuracy. The pass-over class showed relatively consistent performance across different filter sizes; however, larger filters outperformed smaller ones, with the $18 \times 18$ configuration providing one of the highest accuracies. In the coalescence class, smaller filters struggled to capture the subtle features necessary for accurate predictions. Starting from filter size $13 \times 13$, the accuracy improved significantly, with the 18 × 18 filter size achieving balanced and high accuracy. Similarly, the reverse-back class benefited from larger filters, with the $18 \times 18$ filter size again standing out for offering balanced and high accuracy.

The comprehensive analysis of filter sizes underscores the critical role of appropriate filter dimensions in enhancing the CNN model's performance. Smaller filters (e.g., $4 \times 4$, $6 \times 6$, etc.) are insufficient for capturing the central features of droplet collisions due to their limited receptive field and the central positioning of droplets in the images. In contrast, larger filters ($13 \times 13$ and above) demonstrate superior performance by effectively extracting relevant features across the entire image.
Among the larger filters, the $18 \times 18$ configuration stands out as the most balanced and effective choice. It achieves high overall validation accuracy and exhibits consistent, balanced class-wise prediction accuracy. Therefore, based on our analysis, we have decided to use the $18 \times 18$ filter size and this choice ensures optimal feature extraction, stability, and overall predictive performance for droplet collision outcomes.

\subsection{Finalized CNN model architecture}

After extensive analysis and optimization, we have developed a CNN architecture that demonstrates better accuracy and balanced performance in predicting droplet collision outcomes. This model for predicting droplet collision outcomes consists of several key layers: convolutional, activation, pooling, and dense layers. The layer by layer description of the finalized model is provided below. Every layer is thoroughly described to highlight its specific characteristics.
\begin{enumerate}
    \item \textbf{Input layer:}
   The model begins with an input layer configured to accept grayscale images of size $164 \times 164 \times 1$ pixels. Each image is represented in a single channel ($1$ for grayscale).

    \item \textbf{Convolutional layer:}
    Following the input layer is a Conv2D layer with $14$ filters of size $18 \times 18$. This layer applies these filters across the input images without padding ('valid'), resulting in output feature maps of size $147 \times 147 \times 14$. The total number of trainable parameters in this layer is $4,550$.

     \item \textbf{Activation layer:}
    After convolution, a ReLU activation function is applied. ReLU introduces non-linearity, enabling the model to learn complex patterns from the feature maps generated by the convolutional layer.

     \item \textbf{Pooling layer:}
    An AveragePooling2D layer with a pool size of $30 \times 30$ follows the activation layer. This pooling layer reduces the spatial dimensions of the feature maps to $4 \times 4 \times 14$, summarizing the presence of features and reducing computational complexity.

    \item \textbf{Flatten layer:}
    The Flatten layer converts the $4 \times 4 \times 14$ feature maps into a single vector of shape ($224,$), preparing the data for the subsequent fully connected layer.

    \item \textbf{Dense layer:}
    A Dense layer with $3$ units and a Softmax activation function is used as the final layer. This layer produces probabilities for each of the three classes (reverse-back, coalescence, pass-over), with each neuron corresponding to one class.

     \item \textbf{Hyperparameters:} Learning rate - $l_r=0.01$, Optimizer - RMSProp, Weight decay - $10^{-6}$, Batch size - $4$, and Epochs - $1500$.
\end{enumerate}
Table~\ref{tab:cnn_model_configuration} summarizes the model architecture and hyperparameters for quick reference. The best validation accuracy achieved by the developed CNN model after playing with different hyperparameters has been displayed in Table~\ref{tab:cnn_best_accuracy}. This performance marks a significant achievement in our primary goal of accurately predicting collision outcomes based on droplet shape features.

\begin{table}[hbt!]

\centering
	\caption{Summarized configuration of the CNN model} 
    \label{tab:cnn_model_configuration}
\vspace*{2mm}

\begin{tabularx}{\textwidth}{|X|X|X|X|}
\hline
\textbf{Layer Type} & \textbf{Output Shape} & \textbf{Parameters} & \textbf{Description}                                   \\ \hline
Input               & (164, 164, 1)         & -                   & Grayscale images of size $164\times164$ pixels                \\ \hline
Conv2D              & (147, 147, 14)        & 4,550               & 14 filters of size $18\times18$, ReLU activation              \\ \hline
AveragePooling2D    & (4, 4, 14)            & 0                   & Pool size of $30\times30$                                     \\ \hline
Flatten             & (224)                 & 0                   & Flattens the pooled feature maps                       \\ \hline
Dense               & (3)                   & 675                 & Fully connected layer with 3 units, Softmax activation \\ \hline
\end{tabularx}
\end{table}
The robustness of a ML based model refers to its ability to maintain high performance across different datasets and varying conditions. To ensure the robustness and generalizability of our model, it is essential to test its performance on a broader range of simulation conditions beyond the training and validation datasets. In the context of droplet collision predictions, this involves evaluating the model's performance under diverse physical conditions that were not part of the original training set. Such conditions include variations in density ratios, viscosity ratios, initial offset distances, and confinement levels. By assessing the model's robustness, we can determine its applicability and reliability in real-world scenarios where these parameters may differ significantly from the controlled conditions used during training.

\begin{table}[hbt!]

\centering
	\caption{Best validation accuracy achieved by the developed CNN model} 
    \label{tab:cnn_best_accuracy}
\vspace*{2mm}

\begin{tabular}{|c|c|c|c|}
\hline
\textbf{Total accuracy} & \textbf{Rev-accuracy} & \textbf{Coal-accuracy} & \textbf{Pass-accuracy} \\ \hline
\textbf{0.969}          & \textbf{1}            & \textbf{0.945}         & \textbf{0.978}         \\ \hline
\end{tabular}
\end{table}

\subsection{Assessing model robustness}

\begin{figure}[H]
  \centering
  \begin{tabular}{ccc}
    \includegraphics[scale = 0.072]{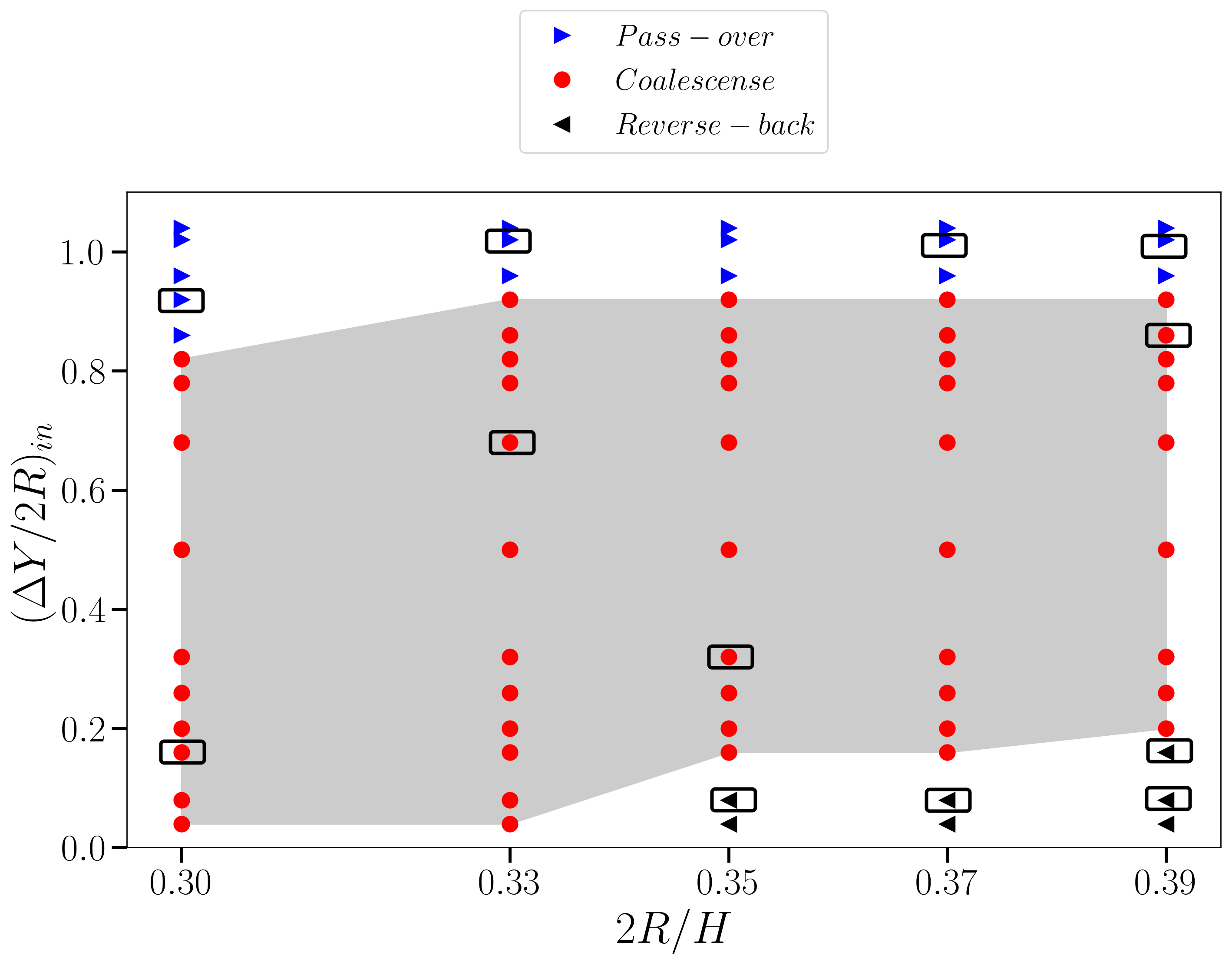} & 
    \includegraphics[scale = 0.072]{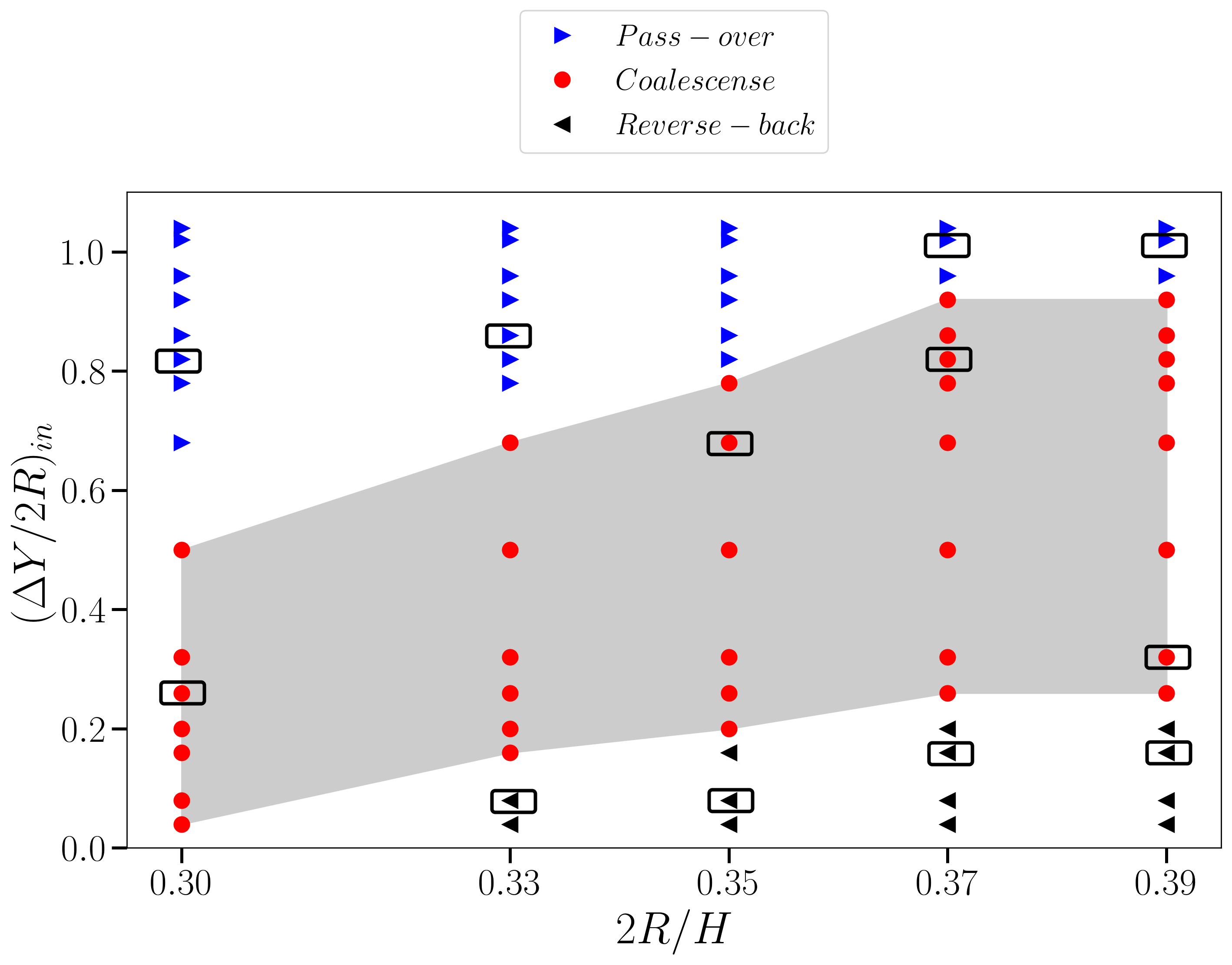} & \\
    (a) $\rho_{12}=60, \mu_{12}=60$ & (b) $\rho_{12}=800, \mu_{12}=60$ & \\
    \multicolumn{2}{c}{\includegraphics[scale = 0.095]{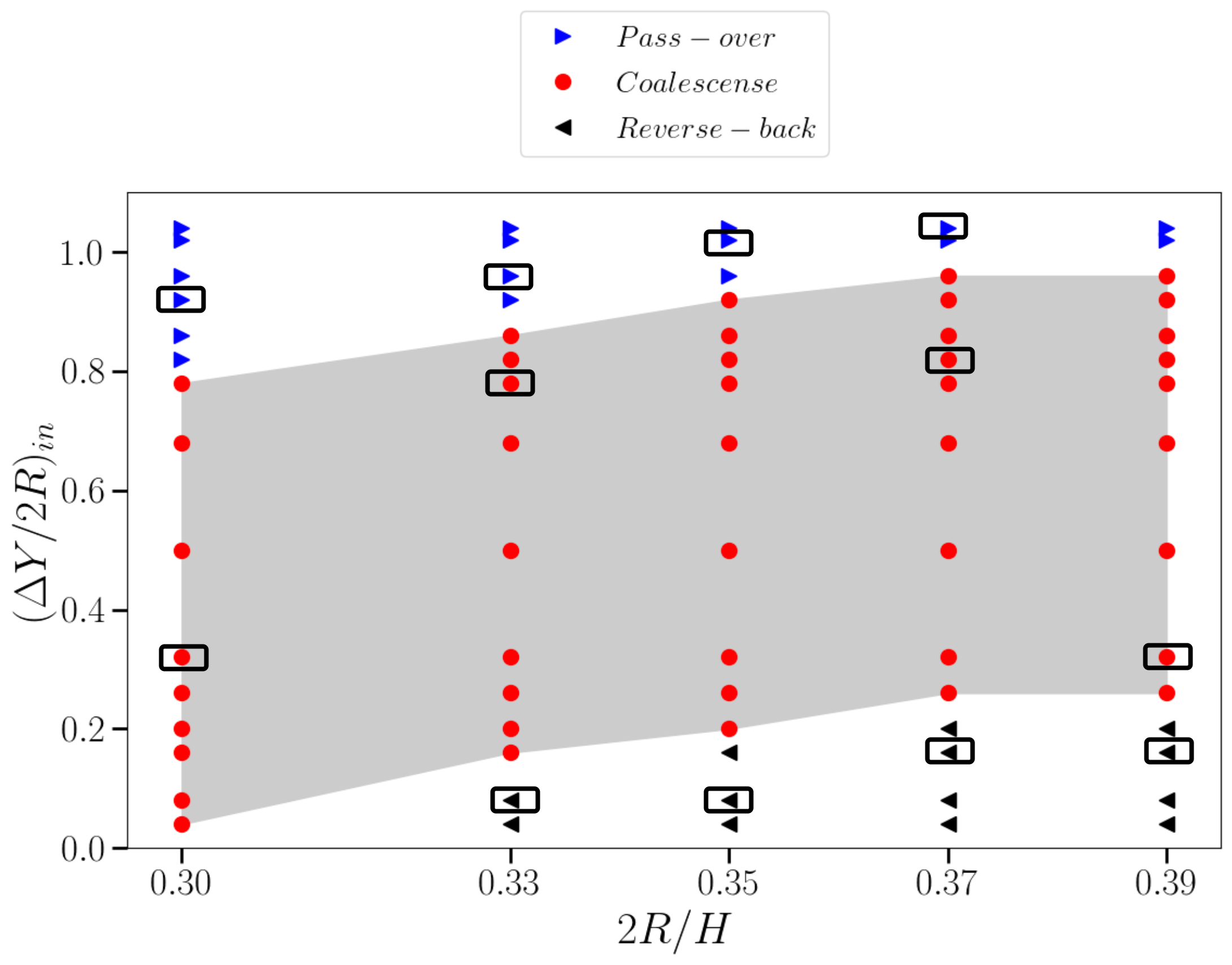}} \\
    \multicolumn{2}{c} {(c) $\rho_{12}=800, \mu_{12}=24$} \\
  \end{tabular}
  \caption{Revised version of Figure~14 from \cite{PhysRevFluids.7.123603}, where black rectangles highlight the 36 cases selected for robustness testing, representing four random samples each from reverse-back, coalescence, and pass-over outcomes for each combination of density and viscosity ratios with $Ca = 0.01$ and $Re = 1.0$.}
  \label{fig:outside_test_cases}
\end{figure}

In our earlier work~\cite{PhysRevFluids.7.123603}, Figure~\ref{fig:outside_test_cases} presents the collision outcomes for varying initial offsets and confinements under three different combinations of density and viscosity ratios that are significantly greater than unity. This figure serves as a critical visual reference for understanding the diverse conditions under which droplet collisions have been analyzed. To thoroughly evaluate the robustness of our developed CNN model, we strategically selected a subset of these cases for further analysis. Our goal is to ensure a comprehensive and balanced assessment of the CNN model's predictive capability across different collision outcomes under varied physical conditions. To this end, we employed the following selection criteria. First, we ensured diversity of collision outcomes by choosing cases representing each of the three collision outcomes: reverse-back, coalescence, and pass-over. This selection allows for evaluating the model’s performance across the spectrum of possible outcomes. Additionally, to capture the effects of varying physical properties, we included cases from each of the three combinations of density and viscosity ratios. These combinations significantly exceed unity, presenting a more challenging test for the model compared to the unit ratio cases used during training. Finally, for balanced assessment, we randomly picked four cases for each collision outcome and each combination of density and viscosity ratios. This random sampling approach mitigates selection bias and ensures a fair representation of the dataset.
From the identified cases, we selected a total of 36 scenarios, 4 reverse-back, 4 coalescence, and 4 pass-over cases for each of the three combinations of density and viscosity ratios. These 36 cases provide a diverse and representative sample of the broader dataset, covering various initial offsets and confinement levels. To facilitate a detailed robustness assessment, we collected 360 snapshots from these 36 selected cases. Each snapshot captures the droplets in close proximity just before collision, offering critical shape information necessary for accurate prediction. This substantial dataset ensures that our analysis is statistically significant and that the model's performance can be rigorously evaluated. In the revised version of Figure-14 from~\cite{PhysRevFluids.7.123603}, as depicted in Figure~\ref{fig:outside_test_cases}, the selected cases are highlighted with black rectangles, providing a clear visual indication of the scenarios chosen for robustness testing. The selected test cases and collected snapshots form the foundation for the robustness check of our CNN model. By applying the model to these diverse scenarios, we aim to ($1$) evaluate predictive accuracy, ($2$) assess generalizability, and ($3$) identify performance trends.

The successful application of our developed CNN model to test cases featuring significantly higher density and viscosity ratios underscores its robustness and generalizability. With a total accuracy of $97.2\%$, our model demonstrates exceptional performance in predicting the outcome of droplet collisions under varied conditions for all the cases as provided in Table~\ref{tab:accuracy_out_test_data}

In terms of total accuracy this model achieved an impressive total accuracy of $97.2\%$, reaffirming its reliability in accurately predicting collision outcomes across diverse scenarios. In predicting the reverse-back outcome, this model achieved perfect accuracy ($100\%$), indicating its proficiency in discerning this specific collision outcome. Additionally, our model exhibited commendable accuracy in predicting coalescence ($94.2\%$) and pass-over ($97.5\%$) outcomes, further validating its effectiveness in capturing the subtle dynamics of droplet collisions.

\begin{table}[hbt!]

\centering
	\caption{Accuracy Achieved on the Outside Test Data} 
    \label{tab:accuracy_out_test_data}
\vspace*{2mm}

\begin{tabular}{|c|c|c|c|}
\hline
\textbf{Total accuracy} & \textbf{Rev-accuracy} & \textbf{Coal-accuracy} & \textbf{Pass-accuracy} \\ \hline
\textbf{0.972}          & \textbf{1.0}            & \textbf{0.942}         & \textbf{0.975}         \\ \hline
\end{tabular}
\end{table}

\section{\label{sec:conclusion}Conclusion}

In this study, we developed and applied a CNN model to predict the outcomes of droplet pair collisions in a microchannel based on droplet shape deformation during the collision process. By leveraging CNN’s ability to capture complex, nonlinear relationships in visual data, we achieved high accuracy in predicting key outcomes such as coalescence, pass-over, and reverse-back events. A lattice Boltzmann method was used to numerically generate dataset for pair droplet collisions in confined shear flow for modeling the network. A key aspect of our work involved the systematic examination of various hyperparameters, including the optimization method, learning rate, filter dimensions, and filter numbers. Through extensive experimentation and analysis, we determined which combinations of these parameters produced the best results. For instance, we found that selecting the appropriate filter size was crucial in capturing the fine details of droplet shape, while the choice of learning rate significantly impacted the model’s convergence and stability during training. The optimization method also played a critical role in improving the model’s performance, balancing both accuracy and computational efficiency. By fine-tuning these hyperparameters, we were able to enhance the model’s predictive capabilities and minimize overfitting, even with a relatively small dataset. To assess the robustness of our model, we tested it on a new dataset that featured different physical properties—specifically, varying density and viscosity ratios compared to the dataset used for training. Despite these differences, the CNN demonstrated strong predictive power, achieving good accuracy ($97.2\%$) in predicting collision outcomes on the new dataset. This result underscores the model’s flexibility and potential to generalize across different scenarios, even when trained on a limited set of conditions. Given that microchannel flow typically falls within the Stokes flow regime, we restricted our analysis to $Re=1$ and  $Ca=0.01$ to capture all relevant outcomes.  While different  $Re$ and $Ca$ values could also be modeled using our approach, this initial range was chosen to balance data generation time with effective modeling. The fact that the CNN performed well with a small dataset shows its potential to scale with more comprehensive data. Incorporating larger datasets, which account for more diverse and detailed physical factors, could further enhance the model's generalization capabilities. This opens up exciting possibilities for future research, where the CNN could be extended to more complex multiphase flow problems involving a wider range of droplet characteristics and collision scenarios. In conclusion, our study demonstrates the feasibility and effectiveness of using CNNs to predict droplet collision outcomes. The promising results achieved with small datasets suggest that, with the inclusion of larger and more varied datasets, the model could be generalized to cover a broader range of flow conditions and collision behaviors. This approach not only reduces the computational burden associated with traditional methods but also provides a scalable framework for analyzing droplet dynamics in microchannels and beyond. \\ \\
\textbf{AUTHOR DECLARATIONS}\\
\textbf{Conflict of Interest}\\
The authors have no conflicts to disclose.\\ \\
\textbf{DATA AVAILABILITY}\\
The data supporting this study’s findings are available from the
corresponding author upon reasonable request.

\pagebreak
\providecommand{\noopsort}[1]{}\providecommand{\singleletter}[1]{#1}%

\end{document}